\documentclass[11pt]{article}
\usepackage{amsfonts,amsbsy,bm,euscript,mathrsfs}
\usepackage{amssymb,stmaryrd,faktor}
\usepackage{amsmath}
\usepackage{mathabx}
\usepackage{bbm}
\usepackage{graphicx}
\usepackage[title,titletoc]{appendix}
\usepackage[bookmarks=true,colorlinks=true,linkcolor=blue,citecolor=blue,urlcolor=blue,bookmarksnumbered]{hyperref}
\usepackage{dsfont}
\usepackage{lmodern}
\usepackage{mathrsfs}
\usepackage{mathtools}
\usepackage{bbm}
\usepackage{braket}
\usepackage{slashed}
\usepackage[normalem]{ulem}
\usepackage{booktabs}
\usepackage{subfig}
\usepackage{tikz}
\usepackage[utf8]{inputenc}
\usepackage[english]{babel}
\usepackage{a4wide}
\usepackage{cite}
\usepackage{appendix}
\usepackage{etoolbox}
\usepackage{physics}
\usepackage{sectsty}
\usepackage{float}
\usepackage{systeme}
\usepackage{xcolor}
\usepackage{mathdots}
\usepackage{nicematrix}
\usepackage{accents}
\usepackage{subcaption}

\graphicspath{{./figures/}}

\DeclareMathOperator{\im}{i\!}
\DeclareMathOperator{\si}{if}

\def
\texto
       {
        \topmargin -20pt
        \oddsidemargin 0pt
        \headheight 0pt
        \headsep 0pt
        \textwidth 6.35in
        \textheight 9.25in
        }
\texto

\def\baselinestretch{1.2}
\sectionfont{\large}
\renewcommand{\theequation}{\thesection.\arabic{equation}}
\csname @addtoreset\endcsname{equation}{section}
\allowdisplaybreaks

\begin{document} 

\begin{titlepage}

\begin{center}

{\large \textbf{Bethe Ansatz, Quantum Circuits, and the F-basis}}

\vskip 0.45in
\textbf{
Roberto Ruiz\footnote{{\texttt{roberto.ruiz@ift.csic.es}}}$^{a}$,
Alejandro Sopena\footnote{{\texttt{alejandro.sopena@tii.ae}}}$^{b,a}$,
Esperanza L{\'o}pez\footnote{{\texttt{esperanza.lopez@csic.es}}}$^{a}$,
Germ{\'a}n Sierra\footnote{{\texttt{german.sierra@csic.es}}}$^{a}$,\\
Bal{\'a}zs Pozsgay\footnote{{\texttt{pozsgay.balazs@ttk.elte.hu}}}$^{c}$
\setcounter{footnote}{0}
}

\vskip 0.15in

$\vphantom{x}^{a}$\! 
Instituto de F{\'i}sica Te{\'o}rica UAM/CSIC \\
Universidad Aut{\'o}noma de Madrid \\
C/ Nicolás Cabrera 13--15 \\
Cantoblanco, 28049 Madrid, Spain \\ 

\vskip 0.15in

$\vphantom{x}^{b}$\! 
Quantum Research Center\\ 
Technology Innovation Institute\\ 
Abu Dhabi, UAE \\

\vskip 0.15in

$\vphantom{x}^{c}$\! 
MTA-ELTE ``Momentum'' Integrable Quantum Dynamics Research Group,\\
ELTE Eötvös Loránd University\\
Pázmány Péter sétány 1/a,\\
1117
Budapest, Hungary \\

\vskip 0.1in

\vskip 0.1in 

\end{center}

\vskip 0.5in

\noindent 
The Bethe Ansatz is a method for constructing exact eigenstates of quantum-integrable spin chains. 
Recently, 
deterministic quantum algorithms, 
referred to as 
``algebraic Bethe circuits", 
have been developed to prepare Bethe states for the spin-$1/2$ XXZ model. 
These circuits represent a unitary formulation of the standard algebraic Bethe Ansatz, 
expressed using matrix-product states that act on both the spin chain and an auxiliary space.
In this work,    
we systematize these previous results, 
and show that algebraic Bethe circuits can be derived by a change of basis in the auxiliary space. 
The new basis, 
identical to the ``F-basis" known from the theory of quantum-integrable models, 
generates the linear superposition of plane waves that is characteristic of the coordinate Bethe Ansatz.
We explain this connection, 
highlighting that certain properties of the F-basis 
(namely, 
the exchange symmetry of the spins) 
are crucial for the construction of algebraic Bethe circuits.
We demonstrate our approach by presenting new quantum circuits for the inhomogeneous spin-$1/2$ XXZ model.

\begin{sloppypar}

\noindent

\end{sloppypar}

\vskip .4in 

\vskip .1in

\noindent

\end{titlepage} 

\def\baselinestretch{1}
\baselineskip 15pt
\sectionfont{\large} 
\renewcommand{\theequation}{\thesection.\arabic{equation}} 
\csname @addtoreset\endcsname{equation}{section}


\tableofcontents

\section{Introduction}

Quantum-integrable models are distinguished many-body systems in one dimension that possess a tower of commuting conserved 
charges~\cite{Caux10}. 
The Bethe Ansatz is a method to solve quantum-integrable models that have particle conservation, 
providing explicit formulae for energies, 
eigenstates, 
scalar products, 
correlation functions, 
etc. 
The coordinate Bethe Ansatz 
(CBA) 
solves the spectral problem by using linear superpositions of plane waves, 
or ``magnons'', 
as trial 
functions~\cite{Bethe31,Gaudin14}. 
The algebraic Bethe 
Ansatz~(ABA) 
systematizes this approach by the R-matrix and the monodromy 
matrix~\cite{Korepin93,Gomez96,Faddeev96}. 
Both methods enable the construction of Bethe states, 
which are eigenstates of the Hamiltonian when their spectral parameters satisfy the Bethe equations. 

The preparation of Bethe states in 
spin-$1/2$ 
chains has great potential in quantum computing.
For instance, 
Bethe states can be used to initialize quantum algorithms of adiabatic~\cite{Wecker15} and real-time~\cite{Kormos16} evolution,
as well as to benchmark quantum devices. 
Recent research focused on the preparation of Bethe states of the paradigm of quantum-integrable 
\mbox{spin-$1/2$} 
chain:
the homogeneous XXZ model~\footnote{
    The ground state of the anti-ferromagnetic spin chain has been approximated by a double-bracket quantum algorithm in~\cite{Robbiati24}.
    Moreover, the homogeneous XXZ model has been considered in connection to the variational quantum eigensolver in~\cite{Nepomechie20,Raveh24ii}
    and shows promise in sampling certain topological invariants~\cite{Crichigno24}.
},
whose Bethe states form a complete basis of the Hilbert space~\cite{Faddeev81}.

A first class of quantum algorithms~\cite{VanDyke21,VanDyke22,Raveh24} 
are based on the special simplicity of Bethe states in the homogeneous XXZ model, 
rather than quantum integrability itself.
They apply in presence of
closed~\cite{VanDyke21,Raveh24} 
and open boundary 
conditions~\cite{VanDyke22,Raveh24},
and are either probabilistic~\cite{VanDyke21,VanDyke22}
or 
deterministic~\cite{Raveh24}.
Algorithms must be efficient to be implementable, 
which,
in the circuit model of quantum computing, 
means that the number of one- and two-qubit gates must grow polynomially with the parameters of the circuit.
The number of gates of probabilistic algorithms is polynomial in the number of qubits 
$N$ 
and magnons 
$M$~\cite{VanDyke21,VanDyke22}. 
However, the success probability decreases exponentially with 
$N$ 
for the ground state, and super-exponentially with
$M$ 
for excited states if $N$ is large \cite{Li22}. 
The number of gates of the deterministic algorithm in~\cite{Raveh24}, 
related to quantum encoders~\cite{Farias24}, 
is linear in
$N$,
but exponential 
in~$M$.

A second class of deterministic algorithms relying explicitly on quantum integrability goes by the name of 
``algebraic Bethe circuits
(ABCs)''~\cite{Sopena22,Ruiz23}.
Just as the ABA builds Bethe states as 
``creation'' 
operators on a 
\mbox{(pseudo-)vacuum},
ABCs seek to frame Bethe states as 
unitaries acting on a reference state.
ABCs were proposed for the homogeneous XXZ 
model with periodic boundaries~\cite{Sopena22,Ruiz23}.
The starting point 
of~\cite{Sopena22} 
was the representation of the Bethe states of the ABA as matrix-product states 
(MPSs)~\cite{Murg12}.
MPSs are the simplest tensor 
networks~\cite{Cirac20}, 
which make the entanglement structure of states 
in one dimension apparent 
by a circuit-like arrangement of local tensors that act on 
both the spin-chain Hilbert space and an auxiliary space.
In~\cite{Sopena22}, the unitaries of ABCs were extracted from the exact tensors of the ABA  
by numerically solving a set of intricate recurrence relations arising from a unitarization procedure.
Closed formulae for the unitaries of ABCs were later obtained 
in~\cite{Ruiz23} 
by a complementary approach.
The key step
in~\cite{Ruiz23} 
was the derivation of an exact representation of
the linear superpositions of plane waves 
of the CBA as an MPS.
The tensors of this MPS,  
unlike those of the ABA,  
directly provide analytical expressions for the unitaries of ABCs,  
as unitarization in this case can be identified  
with the orthonormalization of a basis of Bethe states.
Nonetheless, 
the proof of the construction
of~\cite{Ruiz23} 
was not complete and partially relied on numerical checks for small number of 
magnons~$M$.
The equivalence between the realizations of ABCs in~\cite{Sopena22,Ruiz23}, 
and thus between the formulations of the ABA and the CBA, 
was also verified for small~$M$ in~\cite{Ruiz23}.
The computation of the unitaries of ABCs does not require Bethe states to be eigenstates of the Hamiltonian; 
that is, the Bethe equations need not hold.
Moreover, the number of unitaries of ABCs is linear in the number of qubits 
$N$.
However, the ABC unitaries act on up to
$M+1$~qubits, and the efficiency of their decomposition into one- and two-qubit gates with respect to 
$M$ remains uncertain in general. 
The generalization of the ABC construction 
to open boundaries appeared
in~\cite{Ruiz24}.

\begin{figure}[ht]
    \centering
    \includegraphics[width=0.65\textwidth]{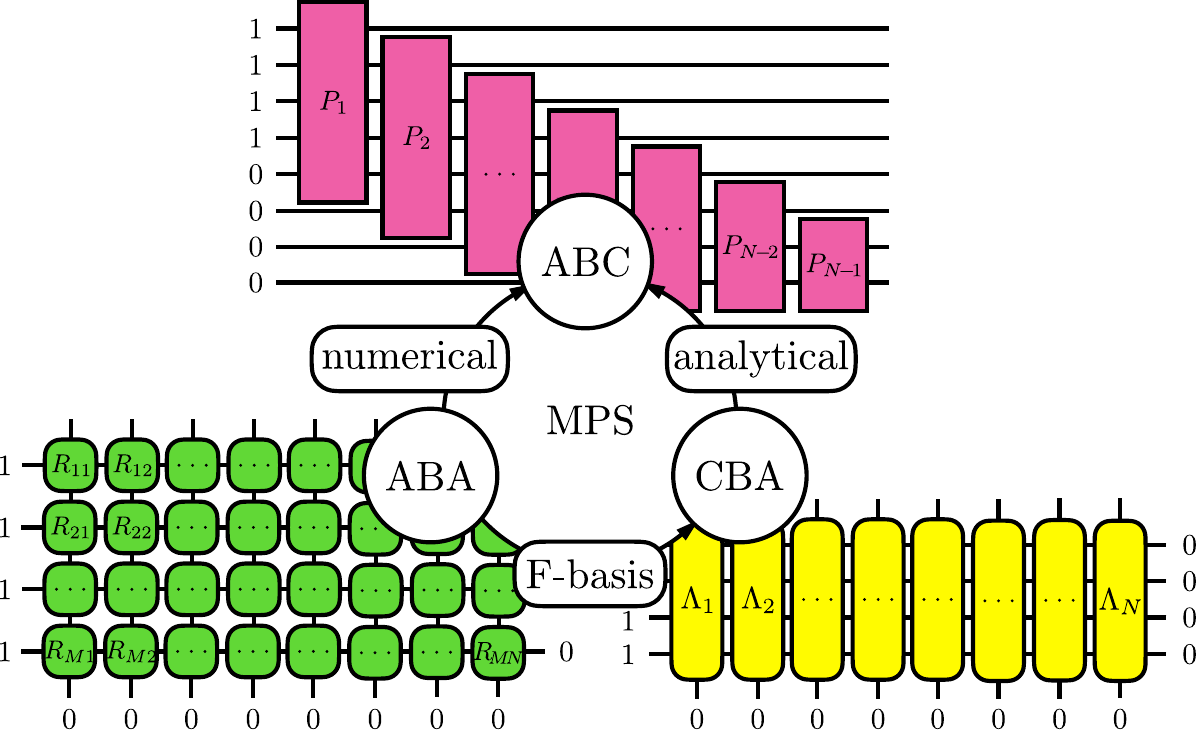}
    \caption{
        Diagram of realizations of the Bethe Ansatz.
    }
    \label{figsummary}
\end{figure}

The results 
of~\cite{Ruiz23} 
raise the question of systematizing ABCs. 
A clear method for formulating Bethe states of the ABA as the circuits 
of~\cite{Ruiz23} 
would enhance the search for integrable models in which the unitaries of ABCs admit efficient decompositions. 
The missing link preventing the systematization of ABCs is the connection between the MPSs of the ABA and the CBA. 
The ABA, 
straightforwardly identifiable with an MPS \cite{Murg12}, 
is the standard method for computing Bethe states, 
while the MPS of the CBA 
provides closed formulae for the unitaries of ABCs. 
In this work, 
we propose that the change to the F-basis 
of~\cite{Maillet96} 
in the auxiliary space is 
the key to transforming the MPS 
of the ABA into that of the CBA,
thereby enabling the analytical 
reformulation of Bethe states as quantum circuits. 
The crucial property of the F-basis 
is its invariance under exchange of qubits,
which characterizes the MPS of the CBA.
We also address the loopholes in~\cite{Ruiz23}  
by presenting a rigorous method 
to eliminate post-selected qubits in the circuit.
We illustrate our approach 
with new ABCs for
the inhomogeneous XXZ model with periodic boundaries.
Figure~\ref{figsummary} summarizes 
the realizations of
the Bethe Ansatz that we uncover.

The F-basis is a special basis where the operators of the ABA are local operators with a 
\mbox{non-local} 
dressing~\cite{Maillet96}. 
Originally proposed for the inhomogeneous XXZ 
model with periodic boundaries~\cite{Maillet96},
the F-basis, 
which admits a diagrammatic 
representation~\cite{McAteer11}, 
was soon extended to encompass
higher-spin~\cite{Terras99,Pfeiffer00,Pfeiffer00ii}, 
higher-rank~\cite{Albert00,Yang05,McAteer12},
totally anisotropic~\cite{Albert00ii}, 
open~\cite{Wang02,Kitanine07}, and
supersymmetric~\cite{Yang04,Zhao05}
chains, among 
others~\cite{Maruyama10,Qiao18}. (We refer to  \cite{Kitanine98,McAteer11} for a summary of the F-basis.)
The F-basis in the quantum space led to the first solution to the quantum inverse scattering problem for the spin 
operators~\cite{Kitanine98}, 
paving the way for the exact 
computation of form factors and correlation functions.
Other applications included 
the computation of domain-wall 
partition functions~\cite{Martin11,Zhong22},
asymmetric simple exclusion 
processes~\cite{Crampe15},
and spin-chain 
propagators~\cite{Feher19}.
The importance of applying the F-basis to the MPS of the Bethe Ansatz was envisaged 
in~\cite{Katsura12}, 
which reviewed the equivalence of the MPS 
of~\cite{Alcaraz03,Alcaraz03ii,Alcaraz06}
with the ABA by a change of basis in the auxiliary space as presented in~\cite{Katsura10}.
The same remark was made in~\cite{Ovchinnikov10}, 
which also underscored
the suitability of the F-basis 
for the explicit computation of Bethe wave functions.

The paper has the following structure.
In 
Section~\ref{sFXXZ}, 
we present the inhomogeneous XXZ model with periodic boundaries, 
the monodromy matrix, 
and the R-matrix.
We introduce the F-basis in the auxiliary space
and show it is symmetric with respect to the exchange of qubits. 
In Section~\ref{sFABC},
we prove the change to the F-basis in the auxiliary space relates the formulation of the ABA and the CBA as MPSs. 
First, 
we focus on the homogeneous XXZ model,
and elaborate on the interplay between the F-basis and Bethe states.
Next, 
we turn to the inhomogeneous XXZ model and derive a simple parameterization of the CBA in this case.
We then use the MPS of the CBA to compute the unitaries of ABCs.
We eliminate post-selected qubits rigorously by a suitably re-defined MPS.
In Section~\ref{sconclusions},
we conclude with general remarks and prospects
on future research.
Appendices~\ref{appMPSCBA}--\ref{appunitaries} provide proofs of the claims in the text and additional material.

\section{The F-basis of the XXZ Model}

\label{sFXXZ}

In this section, 
we review the inhomogeneous
XXZ model with periodic boundaries.
In Subsection~\ref{ssRT}, 
we introduce the model, the R-matrix, and the monodromy matrix.
We present the exchange algebra of the monodromy matrix, 
whereby the ABA follows.
In subsection \ref{ssF}, we review the F-basis of~\cite{Maillet96}.
We highlight that operators of the ABA in the F-basis are symmetric with respect to exchange of qubits.
The property proves to be instrumental
in derivation of
ABCs in
Section \ref{sFABC}
into this analysis.
We refer to \cite{Kitanine98,McAteer11}
for a summary 
of the F-basis of~\cite{Maillet96}.

\subsection{The R-matrix and Monodromy Matrix}

\label{ssRT}

We begin by briefly reviewing the ABA for the homogeneous XXZ model with periodic boundaries. 
The review serves to both provide context and facilitate the extension of the ABA to the inhomogeneous spin chain. 
The XXZ model is a chain of 
$N$ spin-$1/2$ sites, or, alternatively,
``qubits''. 
The Hilbert space is
\begin{equation}
    \label{HN}
    \mathsf{H}_N=
    \overset{N}{\underset{j=1}{\bigotimes}}~\mathsf{h}_{j} \ ,
    \quad \mathsf{h}_{j}\cong\mathbb{C}^{2} \ ,
\end{equation}
and goes by the name of 
``quantum space''. 
We call the qubits of the quantum space
``spins'' 
to differentiate them from auxiliary qubits below.
The subscript 
$j=1,\ldots,N$ 
labels the individual Hilbert space of the spins 
$\mathsf{h}_j$,
which is isomorphic to 
$\mathbb{C}^2$. 
The computational basis of a qubit corresponds to up and down spin-$1/2$ states like
\begin{equation}
    \label{basisvectors}
    \ket{\uparrow} := \ket{0}=
    \begin{bmatrix}
        1 \\
        0
    \end{bmatrix} \ , \quad 
    \ket{\downarrow} := \ket{1}=
    \begin{bmatrix}
        0 \\
        1
    \end{bmatrix} \ .
\end{equation}
The Hamiltonian is
\begin{equation}
    \label{H}
    H=\sum_{j=1}^{N} \left(X_jX_{j+1}+Y_jY_{j+1}+\Delta Z_jZ_{j+1} \right) \ ,
\end{equation}
where 
$\Delta$ 
is the anisotropy parameter.
We introduced standard Pauli matrices on the 
$j$-th 
spin subject to periodicity.
To diagonalize the Hamiltonian,
one considers the monodromy matrix, 
the 
$2\times 2$-matrix 
whose entries are the operators of the ABA on the quantum space:
\begin{equation}
    \label{ABA}
        T(u)=
        \begin{bmatrix}
            A(u) & B(u) \\
            C(u) & D(u)
        \end{bmatrix} 
    \in\mathrm{End}(\mathsf{h}_0\otimes \mathsf{H}_N)\ , 
    \quad \mathsf{h}_0\cong\mathbb{C}^2 \ .
\end{equation}
The variable 
$u$ 
denotes the spectral parameter.
The 
$2\times2$-matrix 
acts on
$\mathsf{h}_0$,
called ``auxiliary space'', 
which corresponds to an auxiliary qubit called 
``ancilla''.
The ABA dictates the construction of eigenstates by applying one of the non-diagonal operators from the monodromy matrix to a reference state,
whose spectral parameters must fulfil the Bethe equations.
These Bethe states not only diagonalize the Hamiltonian, 
but also the transfer matrix,
\begin{equation}
    \label{transfer}
    t(u)=\mathrm{tr}\,T(u)=A(u)+D(u) \ ,
\end{equation}
for every 
$u$. 
The regular series of the transfer matrix around every point spans a tower of commuting conserved charges diagonalized by the ABA,
the standard hallmark of quantum integrability~\cite{Caux10}. 
The cornerstone of the ABA is the R-matrix,
which we introduce next.

\begin{figure}[ht]
    \centering
    \includegraphics[width=0.275\textwidth]{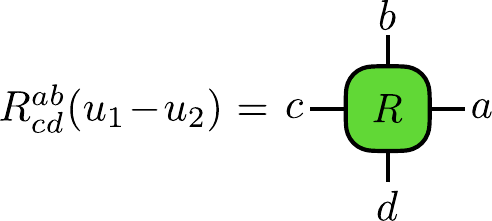}
    \caption{
        The R-matrix (\ref{Rmatrix}) as a two-qubit tensor.
        The R-matrix acts from left to right on the first qubit, associated to $u_1$,
        and from bottom to top on the second, associated to $u_2$.
    }
    \label{figR}
\end{figure}

The R-matrix of the XXZ model is
\begin{equation}
    \label{Rmatrix}
    R(u) = 
    \begin{bmatrix}
        1 & 0 & 0 & 0 \\
        0 & f(u) & g(u) & 0 \\
        0 & g(u) & f(u) & 0 \\
        0 & 0 & 0 & 1
    \end{bmatrix}
     \ , \quad f(u)=\frac{\sinh u}{\sinh(u+\im\gamma)} \ , \quad g(u)=\frac{\sinh(\im\gamma)}{\sinh(u+\im\gamma)} \ .
\end{equation}
The anisotropy parameter
$\Delta$ 
that characterizes the spin chain depends on 
$\gamma$ 
like~\footnote{
    References~\cite{Sopena22,Ruiz23} 
    mainly addresed the critical homogeneous XXZ model,
    where $-1<\Delta\leq 1$ ($0\leq \gamma<\pi$).
    Since we also consider the inhomogeneous spin chain here,
    whose phase diagram is not known, 
    we 
    allow complex~$\gamma$.
}
\begin{equation}
    \Delta=\cos\gamma \ .
\end{equation} 
The R-matrix is an operator on the Hilbert space of two qubits, and
we understand $R(u)$ as a two-qubit tensor which depends on the difference of the spectral parameters of each 
qubit~$u:=u_1-u_2$. 
The non-vanishing components of the R-matrix are
\begin{equation}
    R_{00}^{00}(u)=R_{11}^{11}(u)=1 \ , \quad R_{01}^{01}(u)=R_{10}^{10}(u)=f(u) \ , \quad R_{01}^{10}(u)=R_{10}^{01}(u)=g(u) \ .
\end{equation}
We depict the R-matrix as a tensor in Figure \ref{figR}.

\begin{figure}[ht]
    \centering
    \includegraphics[width=0.3\textwidth]{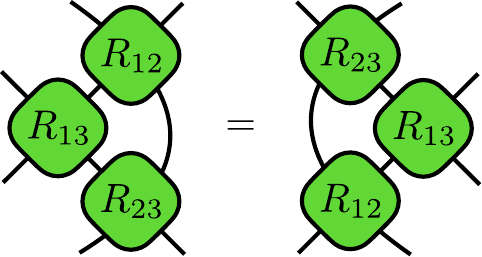}
    \caption{
        The YBE (\ref{YBdiff}) as an equality of tensor networks.
        Each R-matrix acts on two qubits from bottom to top,
        and so it is read the diagram.
    }
    \label{figYB}
\end{figure}

The R-matrix satisfies the Yang-Baxter equation
(YBE) in difference form:
\begin{equation}
    \label{YBdiff}
    R_{12}(u_1-u_2)R_{13}(u_1-u_3)R_{23}(u_2-u_3) = R_{23}(u_2-u_3)R_{13}(u_1-u_3)R_{12}(u_1-u_2) \ .
\end{equation}
Subscripts denote the Hilbert space of the pair of qubits the R-matrix acts on: 
\begin{equation}
    R_{12}=R\otimes 1_2 \ , \quad R_{23}= 1_2\otimes R \ , \quad 
    R_{13}=(\Pi\otimes  1_2)\,R_{23}\,(\Pi\otimes   1_2)\ ,
\end{equation}
where 
$1_2$ 
denotes the identity 
$2\times 2$-matrix 
and 
the transposition 
$4\times 4$-matrix $\Pi$ is
\begin{equation}
    \label{Pi}
    \Pi=
    \begin{bmatrix}
        1 & 0 & 0 & 0 \\
        0 & 0 & 1 & 0 \\
        0 & 1 & 0 & 0 \\
        0 & 0 & 0 & 1
    \end{bmatrix} \ .
\end{equation}
Figure~\ref{figYB} 
depicts the YBE as an equality between tensor networks of R-matrices.

\begin{figure}[ht]
    \centering
    \includegraphics[width=\textwidth]{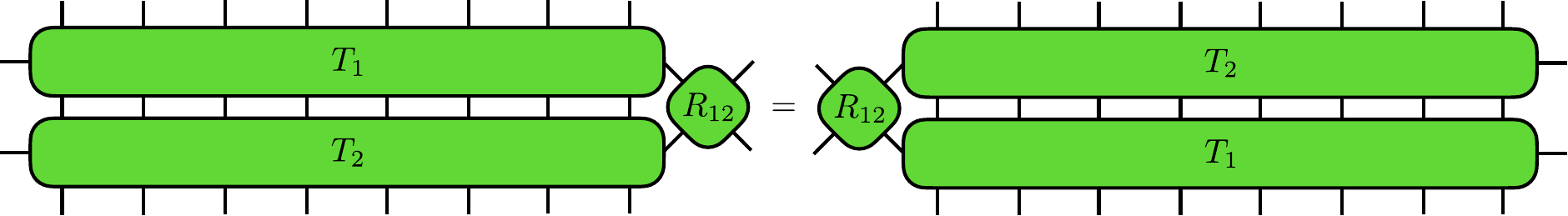}
    \caption{
    The RTT-relation 
    (\ref{RTT}) 
    via tensor networks.
    The monodromy and R-matrices act on ancillae from left to right.
    The monodromy matrices act on spins from bottom to top.
    }
    \label{figRTT}
\end{figure}

The monodromy matrix (\ref{ABA}) spans an algebra with respect to the product in the auxiliary space called
``exchange algebra''. 
The algebra is associative and unital, but not commutative.
The R-matrix is the intertwiner that encodes non-commutativity through the RTT-relation:
\begin{equation}
    \label{RTT}
    R_{12}(u-v)T_1(u)T_2(v) = T_2(v)T_1(u)R_{12}(u-v) \ ,
\end{equation}
where
\begin{equation}
    T_1=T\otimes 1_2 \ , \quad T_2=1_2\otimes T \ .
\end{equation}
Note the R-matrix acts on the auxiliary space of two ancillae,
and the monodromy matrix acts on the Hilbert space of an ancilla and on the quantum space.
Figure~\ref{figRTT} 
depicts the RTT-relation as an equality between tensors.
The RTT-relation
specifies the standard commutation relations of the ABA of the XXZ model; see 
(1.11)–(1.24)
of Chapter VII of \cite{Korepin93}.
The RTT-relation also implies that transfer matrices with different spectral parameters commute,
hence that they are simultaneously diagonalizable.

\begin{figure}[ht]
    \centering
    \includegraphics[width=0.9\textwidth]{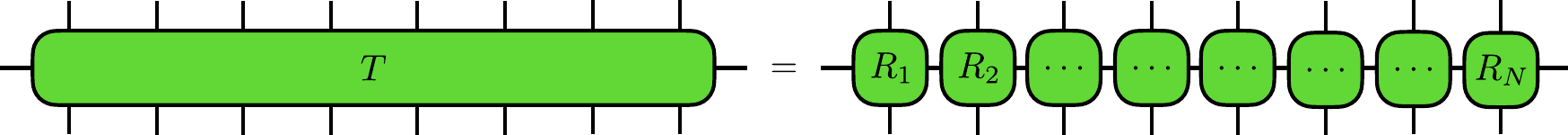}
    \caption{The monodromy matrix (\ref{TXXZ}) as a tensor network. The notation 
    is~\mbox{$R_{j}:=R_{0j}(u-v_j)$}. 
    The action on ancillae goes from left to right and on spins from bottom to top.
    The R-matrices on the diagram and the equation appear in reverse order.
    In the diagram, the leftmost $R_1$ acts first, 
    the next-to-leftmost $R_2$ acts second, etc.
    In the equation, the rightmost $R_1$ acts first, 
    the next-to-rightmost $R_2$ acts second, etc.
    }
    \label{figT}
\end{figure}

The monodromy matrix of the XXZ model is a tensor network of R-matrices, 
where the YBE~\mbox{(\ref{YBdiff})}
implies the RTT-relation
(\ref{RTT}).
Let 
$R_{0j}\in\mathrm{End}
(\mathsf{h}_0\otimes\mathsf{h}_j)$
be the R-matrix of the
ancilla and the $j$-th spin.
The monodromy matrix of the inhomogeneous XXZ model is
\begin{equation}
    \label{TXXZ}
    T(u)=R_{0N}(u-v_N)\ldots R_{02}(u-v_2) R_{01}(u-v_1) \ ,
\end{equation}
where 
$v_j$ 
is the inhomogeneity of the 
$j$-th spin.
Figure 
\ref{figT} 
illustrates the monodromy matrix as a tensor network of R-matrices.
The YBE 
(\ref{YBdiff})
implies the RTT-relation 
(\ref{RTT}).
Even though
(\ref{TXXZ})
defines a one-parameter family of transfer matrices by
(\ref{transfer}),
inhomogeneities forbid the tower of conserved charges to be local,
as often required to quantum-integrable models~\cite{Caux10}. 
Lacking a local Hamiltonian, 
we define the inhomogeneous spin chain by the transfer matrix.
If $v_j=v$, 
the spin chain is homogeneous,
and there is a tower of local conserved charges that contains the Hamiltonian~(\ref{H}).
The charges are proportional to logarithmic derivates of the transfer matrix thanks to regularity,
namely,
$R(0)=\Pi$~\cite{Korepin93}.

We close this subsection by noting that the exchange algebra must allow for products of more than two monodromy matrices.
Consider the product of three monodromy matrices.
Inverting the product 
by alternative sequences of pair-wise swappings leads us to  
\begin{equation}
    \label{3T}
    T_1^{\vphantom{-1}}T_2^{\vphantom{-1}}T_3^{\vphantom{-1}}=^{\vphantom{-1}}R_{23}^{-1}R_{13}^{-1}R_{12}^{-1}T_3^{\vphantom{-1}}
    T_2^{\vphantom{-1}}T_1^{\vphantom{-1}}R_{12}^{\vphantom{-1}}R_{13}^{\vphantom{-1}}R_{23}^{\vphantom{-1}}
    =R_{12}^{-1}R_{13}^{-1}R_{23}^{-1}T_3^{\vphantom{-1}}T_2^{\vphantom{-1}}T_1^{\vphantom{-1}}R_{23}^{\vphantom{-1}}
    R_{13}^{\vphantom{-1}}R_{12}^{\vphantom{-1}} \ , 
\end{equation}
where
\begin{equation}
    \label{notationT}
    R_{ab}:=R_{ab}(u_a-u_b) \ , \quad T_a:=T_a(u_a) \ .
\end{equation}
The RTT-relation is compatible with 
(\ref{3T}) thanks to the YBE 
(\ref{YBdiff}),
in the sense that alternative sequences of transpositions giving the same product are equivalent.
The YBE similarly ensures the consistency of products of a higher number of monodromy matrices.

\begin{figure}[ht]
    \centering
    \includegraphics[width=\textwidth]{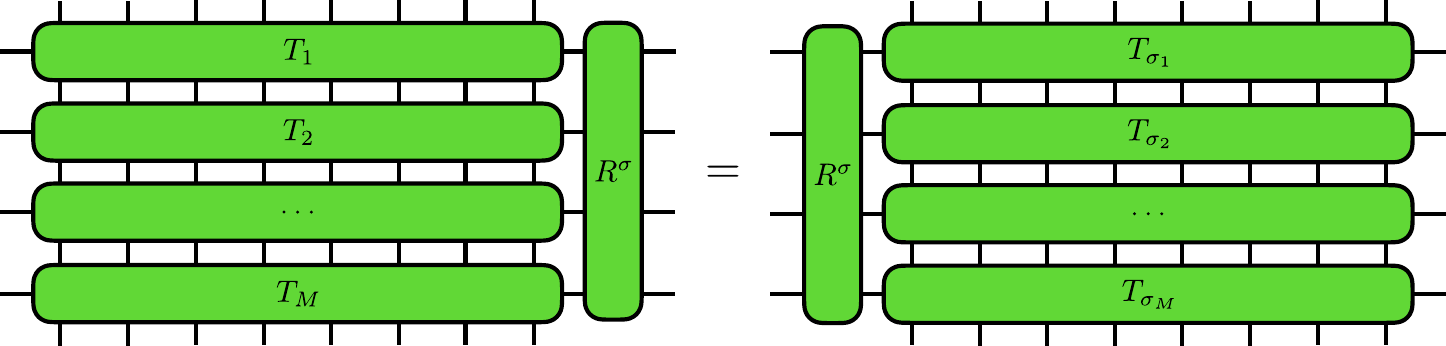}
    \caption{Permutation of the product of $M$ monodromy matrices by $R^{\sigma}:=R^{\sigma}_{12\ldots M}$ in (\ref{LR}).}
    \label{figRTTM}
\end{figure}

The product of two monodromy matrices admits two reorderings,
the trivial reordering and the transposition,
each corresponding to
an element of $S_2$.
Conjugation by $1_{2}$ realizes the identity,
which leaves the order unaffected.
Conjugation by the R-matrix realizes the transposition,
which inverts the order, in agreement with the 
RTT-relation~\mbox{(\ref{RTT})}, 

The product of $M$ monodromy matrices admits $M!$ reorderings.
Reorderings are in \mbox{one-to-one} correspondence with permutations $\sigma\in S_M$. 
Each $\sigma$ corresponds to a $2^M\times 2^M$-matrix $R_{12\ldots M}^{\sigma}$ on the 
auxiliary space of $M$ ancillae
\begin{equation}
    \label{HM}
    \mathsf{H}_{M}=\overset{M}{\underset{a=1}{\bigotimes}}~\mathsf{h}_a \ , 
    \quad \mathsf{h}_a\cong\mathbb{C}^2 \ ,
\end{equation}
which satisfies
\begin{equation}
    \label{LR}
    R_{12\ldots M}^{\sigma}T_1^{\vphantom{\sigma}}T_2^{\vphantom{\sigma}}\ldots 
    T_M^{\vphantom{\sigma}}
    = T_{\sigma_1}^{\vphantom{\sigma}}T_{\sigma_2}^{\vphantom{\sigma}}\ldots 
    T_{\sigma_M}^{\vphantom{\sigma}}
    R_{12\ldots M}^{\sigma} \ ,
\end{equation}
where
\begin{equation}
    \label{notationR}
    R_{12\ldots M}^{\sigma}:=R_{12\ldots M}^{\sigma}(u_1,u_2,\ldots,u_M) \ ,
\end{equation}
We depict (\ref{LR}) in Figure \ref{figRTTM}. Each 
$R_{12\ldots M}^{\sigma}$ 
factorizes into
products of standard R-matrices. 
For instance, 
if $M=3$, 
the 
R-matrix
that performs the
inversion (using cycle notation) $\sigma=(1,3)(2)$ in (\ref{3T}) is
\begin{equation}
    R_{123}^{(1,3)(2)}=R_{12}^{\vphantom{(1)}}R_{13}^{\vphantom{(1)}}R_{23}^{\vphantom{(1)}} \ .
\end{equation}
Alternative factorizations of the same R-matrix are consistent due to the YBE (\ref{YBdiff}).
We are now in position to introduce F-matrices and the F-basis.

\subsection{The F-basis}

\label{ssF}

By definition, 
the F-matrix is the invertible 
$4\times4$-matrix 
that satisfies
\begin{equation}
    \label{RFF}
    R_{12}^{\vphantom{-1}}(u)=F^{-1}_{21}(-u) F_{12}^{\vphantom{-1}}(u)\ ,
\end{equation}
where 
\begin{equation}
    F_{21}=\Pi \, F_{12}\,\Pi \ .
\end{equation}
The F-matrix of the R-matrix
(\ref{Rmatrix}) is
\begin{equation}
    \label{Fmatrix}
    F_{12}(u)=
    \begin{bmatrix}
        1 & 0 & 0 & 0 \\
        0 & 1 & 0 & 0 \\
        0 & g(u) & f(u) & 0 \\
        0 & 0 & 0 & 1 
    \end{bmatrix}
    \ .
\end{equation}
The definition holds thanks to the following properties of the functions in (\ref{Rmatrix}):
\begin{equation}
    f(u)f(-u)+g(u)g(-u)=1 \ , \quad f(u)g(-u)+g(u)f(-u)=0 \ .
\end{equation}
The F-matrix exists because the R-matrix is braided-unitary~\cite{Maillet96}~\footnote{
    Braided unitarity of the R-matrix differs from matrix unitarity in general.
    For instance, (\ref{Rmatrix}) is braided-unitary according to (\ref{unitarityR}),
    but unitary only if $u$ is real.
    We emphasize the name ``unitarity'' for the braided unitarity of the R-matrix (\ref{unitarityR}),
    which 
    borrows from factorized-scattering theory \cite{Zamolodchikov79},
    is deeply ingrained in the literature.
    }:
\begin{equation}
    \label{unitarityR}
    R_{12}(u)R_{21}(-u)=1_{4} \ .
\end{equation}
An F-matrix 
$F_{12\ldots M}$ is the 
$2^M\times 2^M$-matrix that
encodes all the
R-matrices for products of 
$M$ 
monodromy matrices. 
The definition of F-matrices in this case is
\begin{equation}
    \label{RFM}
    R_{12\ldots M}^{\sigma\vphantom{-1}}=
    F_{\sigma_1\sigma_2\ldots\sigma_M}^{-1}F_{12\ldots M}^{\vphantom{-1}} \ ,
\end{equation}
where 
\begin{equation}
    \begin{split}
        F_{12\ldots M}:=F_{12\ldots M}(u_1,u_2,\ldots,u_M) \ , \quad
        F_{\sigma_1\sigma_2\ldots\sigma_M}:=\Pi^\sigma F_{12\ldots M}(u_{\sigma_1},u_{\sigma_2},\ldots,u_{\sigma_M}) \Pi^\sigma \ ,
    \end{split}
\end{equation}
and $\Pi^\sigma$ is the permutation $2^M\times 2^M$-matrix of $\sigma$.
Note that the definition of $F_{\sigma_1\sigma_2\ldots\sigma_M}$ involves both the permutation of ancillae by~$\Pi^\sigma$ and the permutation of the arguments of the F-matrix.
The closed formula of $F_{12\ldots M}$ is~\cite{McAteer11,McAteer12}
\begin{equation}
    \begin{split}
        \label{formFM}
        F_{12\ldots M}
        =
        \overset{M-2}{\underset{a=1}{\prod}}\,\left[\ket{0}_a\bra{0}_a
        +\ket{1}_a\bra{1}_a\overset{M}{\underset{b=a+1}{\prod}}\,
        R_{ab}\right]\ ,
    \end{split}
\end{equation}
where $\ket{i}_a\bra{j}_a$
are projectors
of the Hilbert 
space of the 
$a$-th 
ancilla
and 
\begin{equation}
    R_{ab}:=R_{ab}(u_a-u_b) \ .
\end{equation}
The F-matrices  
$F_{12\ldots M}$ 
realize the reordering of the product of monodromy matrices by means of twists.
The definition 
(\ref{RFM}) 
enables us to rephrase
(\ref{LR}) 
as
\begin{equation}
    \label{FMT}
    F_{12\ldots M}^{\vphantom{-1}}T_1^{\vphantom{-1}}T_2^{\vphantom{-1}}\ldots T_M^{\vphantom{-1}}F_{12\ldots M}^{-1}
    =F_{\sigma_1\sigma_2\ldots\sigma_M}^{\vphantom{-1}}T_{\sigma_1}^{\vphantom{-1}}T_{\sigma_2}^{\vphantom{-1}}\ldots 
    T_{\sigma_M}^{\vphantom{-1}}
    F_{\sigma_1\sigma_2\ldots \sigma_M}^{-1} \ .
\end{equation}
The consistency of $F_{12}$
with the exchange algebra of the monodromy matrix 
implies the consistency among $F_{12\ldots M}$
with different $M$~\cite{Maillet96}.
The F-matrices realize the change
to the ``\mbox{F-basis}'' of the auxiliary space (\ref{HM}), whereby the 
operators of the ABA are
particularly simple~\cite{Maillet96}. 

Before proceeding, we should make a remark. 
Reference 
\cite{Maillet96}
initially introduced F-matrices on the quantum 
space~(\ref{HN}).
The corresponding F-basis is useful for computing scalar products~\mbox{\cite{Maillet96,Kitanine98}}
and solving the quantum inverse scattering problem for local spin operators
\cite{Kitanine98}.
In this work,
however,
we focus on F-matrices on the auxiliary space, 
as proposed 
in \cite{Feher19}.
This approach shall prove to be well-suited for framing Bethe states as quantum circuits.

\begin{figure}[ht]
    \centering
    \includegraphics[width=0.9\textwidth]{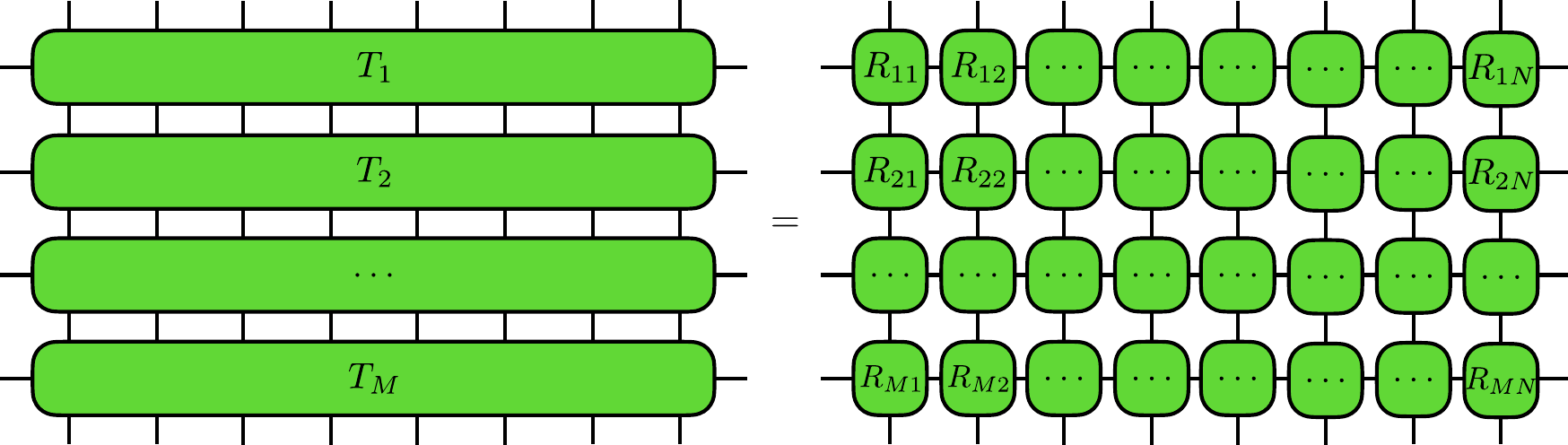}
    \caption{The product of monodromy matrices equals a tensor network of R-matrices.}
    \label{figMT}
\end{figure}

\begin{figure}[ht]
    \centering
    \includegraphics[width=0.9\textwidth]{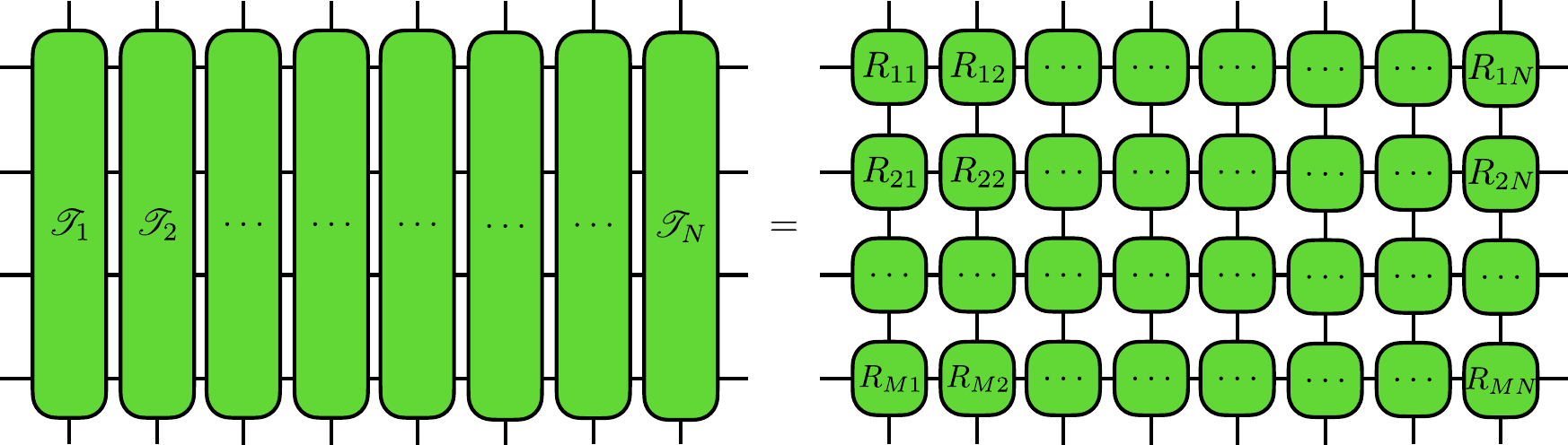}
    \caption{The product of dual monodromy matrices
    equals a tensor network of R-matrices. 
    Grouping R-matrices by columns in
    Figure \ref{figMT}
    leads to this picture.}
    \label{figMTdual}
\end{figure}

Let us consider the product $M$ monodromy matrices.
Figure~\ref{figMT} depicts
the product 
as a tensor network of R-matrices.
Monodromy matrices admit a dual picture where 
spins and ancillae switch roles. 
(We keep the nomenclature ``spins''
and ``ancillae''  in the dual picture with the same meaning
to make the context
clear.)
The definition of the $j$-th
dual monodromy matrix over $M$
ancillae is~\footnote{
    $\mathscr{T}_j$ are often called ``column-to-column'' monodromy matrices,
    as opposed to the ``row-to-row'' matrices $T_a$.
    }
\begin{equation}
    \label{FMTdual}
    \begin{split}
        \mathscr{T}_j(v_j):=\mathscr{T}_j:=R_{1j}(u_1-v_j)
        R_{2j}(u_2-v_j)
        \ldots
        R_{Mj}(u_M-v_j)\ .
    \end{split}
\end{equation}
The spectral parameter of the $j$-th matrix is $v_j$,
whereas $u_a$ is the inhomogeneity of the $a$-th ancilla.
The YBE (\ref{YBdiff}) implies the RTT-relation of dual monodromy matrices is
\begin{equation}
    \label{RTTdual}
    \mathscr{T}_1(v_1)\mathscr{T}_2(v_2)R_{12}(v_1-v_2)=
    R_{12}(v_1-v_2)\mathscr{T}_2(v_2)\mathscr{T}_1(v_1) \ .
\end{equation}
The operators of the ABA of dual monodromy matrices follow from the corresponding exchange algebra.
Figure \ref{figMTdual} depicts
the product of dual monodromy matrices
as a tensor network.

\begin{figure}[ht]
    \centering
    \includegraphics[width=0.4\textwidth]{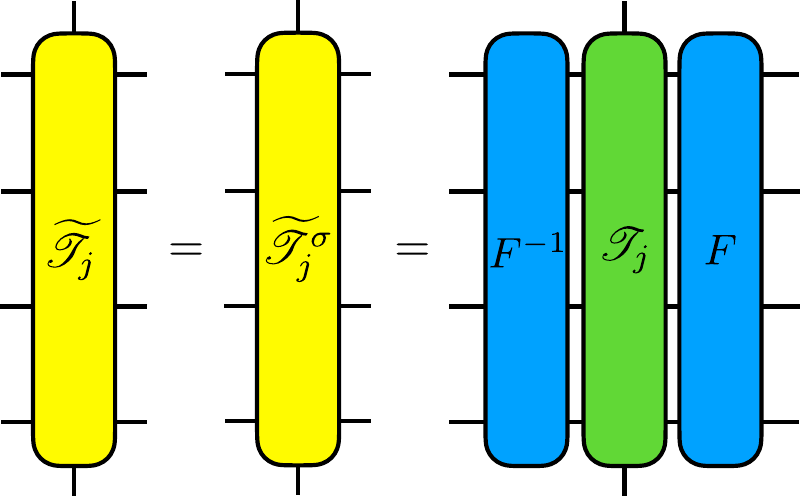}
    \caption{
    New dual monodromy matrices defined in (\ref{symTdual}). 
    The notation is $F=F_{12\ldots M}$.
    These monodromy matrices are symmetric with respect to exchange of ancillae as stated in (\ref{symTdualprop}).
    }
    \label{figsymTdual}
\end{figure}

To introduce the F-basis,
we must perform a change of basis in the auxiliary space by the F-matrix
$F_{12\ldots M}$. 
According to 
(\ref{FMT}), 
the product of monodromy matrices becomes symmetric with respect to permutations once it is twisted by the F-matrix.
This property implies the existence of new dual monodromy matrices
that are symmetric with respect to the exchange of ancillae~\footnote{
    The symmetry of monodromy matrices with respect to the exchange of spins was previously highlighted in \cite{Ovchinnikov00}. We are grateful to A. A. Ovchinnikov for bringing this work to our attention.
}.
Explicitly, the new dual monodromy matrices are
    \begin{equation}
        \label{symTdual}
        \widetilde{\mathscr{T}}_k=F_{12\ldots M}^{\vphantom{-1}}\mathscr{T}_k F^{-1}_{12\ldots M} \ ,
    \end{equation}
satisfying
\begin{equation}
    \label{symTdualprop}
    \widetilde{\mathscr{T}}_k = \widetilde{\mathscr{T}}_k^\sigma \ ,
\end{equation}
where
\begin{equation}
    \widetilde{\mathscr{T}}_k =\widetilde{\mathscr{T}}_k(v_k;u_1,\ldots,u_M) \ ,
    \quad \widetilde{\mathscr{T}}_k^\sigma=\Pi^{\sigma}\widetilde{\mathscr{T}}_k(v_k;u_{\sigma_1},\ldots,u_{\sigma_M})\Pi^{\sigma} \ ,
\end{equation}
and we wrote the dependence on $u_a$ explicitly.
Figure 
\ref{figsymTdual} 
represents 
(\ref{symTdual}) 
and 
(\ref{symTdualprop}). 

The proof of (\ref{symTdualprop})
follows from
the independence 
of the F-matrix
on the number of 
spins~$N$. 
If $N=1$, just
$\widetilde{\mathscr{T}}_1$ 
is defined.
Equation
(\ref{FMT})
with $N=1$
implies
$\widetilde{\mathscr{T}}_1$
fulfills
(\ref{symTdualprop}).
Any other~$\widetilde{\mathscr{T}}_j$ 
is symmetric with respect to exchange
of ancillae because this relation is independent of
$v_j$.
Moreover, we emphasize
$\widetilde{\mathscr{T}}_j$ 
are also dual monodromy matrices.
Since $F_{12\ldots M}$ just acts on the auxiliary space and
does not depend on 
$v_j$, the RTT-relation
(\ref{RTTdual}) holds for $\widetilde{\mathscr{T}}_j$ as well, and
both~$\mathscr{T}_j$ and $\widetilde{\mathscr{T}}_j$ span
the same exchange algebra.
The definition of $\widetilde{\mathscr{T}}_j$ in (\ref{symTdual})
can be understood as a
change of basis
of the dual monodromy 
matrices $\mathscr{T}_j$. 
The new basis is called ``F-basis''.

The operators of the ABA simplify in the F-basis.
Let us write the new dual monodromy matrices as 
$\widetilde{\mathscr{T}}_j\in\mathrm{End}(\mathsf{h}_j\otimes\mathsf{H}_M)$ to mimic (\ref{ABA}).
Then
\begin{equation}
    \label{ABAdual}
    \widetilde{\mathscr{T}}_j=
    \begin{bmatrix}
        \widetilde{\mathscr{A}}_j & \widetilde{\mathscr{B}}_j \\
        \widetilde{\mathscr{C}}_j & \widetilde{\mathscr{D}}_j
    \end{bmatrix} \ .
\end{equation}
The first three operators read~\cite{Maillet96}
\begin{equation}
    \begin{split}
    \label{ABAFbasis}
    \widetilde{\mathscr{A}}_j&=
    \overset{M}{\underset{a=1}{\bigotimes}}
    \begin{bmatrix}
        1 & 0 \\
        0 & f_{aj}
    \end{bmatrix}
    \ , \\
    \widetilde{\mathscr{B}}_j&=
    \sum_{a=1}^{M}
    \overset{a-1}{\underset{b=1}{\bigotimes}}
    \begin{bmatrix}
        1 & 0 \\
        0 & f_{bj}/f_{ba}
    \end{bmatrix}
    \begin{bmatrix}
        0 & 0 \\
        g_{aj} & 0
    \end{bmatrix}
    \overset{M}{\underset{c=a+1}{\bigotimes}}
    \begin{bmatrix}
        1 & 0 \\
        0 & f_{cj}/f_{ca}
    \end{bmatrix}
    \ ,  \\
    \widetilde{\mathscr{C}}_j&=
    \sum_{a=1}^{M}
    \overset{a-1}{\underset{b=1}{\bigotimes}}
    \begin{bmatrix}
        1/f_{ab} & 0 \\
        0 & f_{bj}
    \end{bmatrix}
    \begin{bmatrix}
        0 & g_{aj} \\
        0 & 0
    \end{bmatrix}
    \overset{M}{\underset{c=a+1}{\bigotimes}}
    \begin{bmatrix}
        1/f_{ac} & 0 \\
        0 & f_{cj}
    \end{bmatrix}
    \ ,
    \end{split}
\end{equation}
with 
\begin{equation}
    \begin{split}
        f_{aj}:=f(u_a-v_j) \ , \quad f_{ab}:=f(u_a-u_b) \ , \quad g_{aj}:=g(u_a-v_j)  \ .
    \end{split}
\end{equation}
The operator $\widetilde{\mathscr{D}}_j$
follows from the proportionality of the quantum
determinant of $\widetilde{\mathscr{T}}_j$
to the identity matrix~\cite{Korepin93}.

\section{The F-basis and Algebraic Bethe Circuits}

\label{sFABC}

In this section, we present
the ABCs
of~\cite{Sopena22,Ruiz23} 
in light of the F-basis
of~\cite{Maillet96}. 
Inspired by \cite{Ruiz23},
we use the F-basis, 
starting from the ABA, 
to express the linear superposition of plane waves of the CBA as an MPS.
The tensors of this one-dimensional network inherit the symmetry with respect to exchange of qubits of the F-basis,
which plays a key role in computing analytic expressions of the unitaries of ABCs.
In Subsection~\ref{ssCBAhom}, 
we relate the MPS formulation of the CBA for the 
homogeneous spin chain
in \cite{Ruiz23}
with the F-basis.
In Subsection \ref{ssCBAinh}, 
we use the connection to derive the CBA of the 
inhomogeneous XXZ model.
In Subsection \ref{ssABCinh},
we construct the ABCs for the inhomogeneous 
spin chain along the lines of \cite{Ruiz23}.

\subsection{The F-basis and Coordinate Bethe Ansatz: Homogeneous Spin Chain}

\label{ssCBAhom}

The ABCs consist of unitaries over various qubits
that act on a reference state
to prepare normalized Bethe states.
References \cite{Sopena22,Ruiz23}
obtained ABCs for the homogeneous
XXZ model, whose Bethe Ansatz is well-known.
The starting point of~\cite{Sopena22}
was the formulation of the ABA as an 
MPS~\cite{Katsura10,Murg12}, 
whereas~\cite{Ruiz23}
began with a new MPS representation
of the CBA.
Both classes of MPSs are connected by a change of basis in the auxiliary space, 
which we show to correspond, essentially, to the change to the F-basis.

\begin{figure}[ht]
    \centering
    \includegraphics[width=\textwidth]{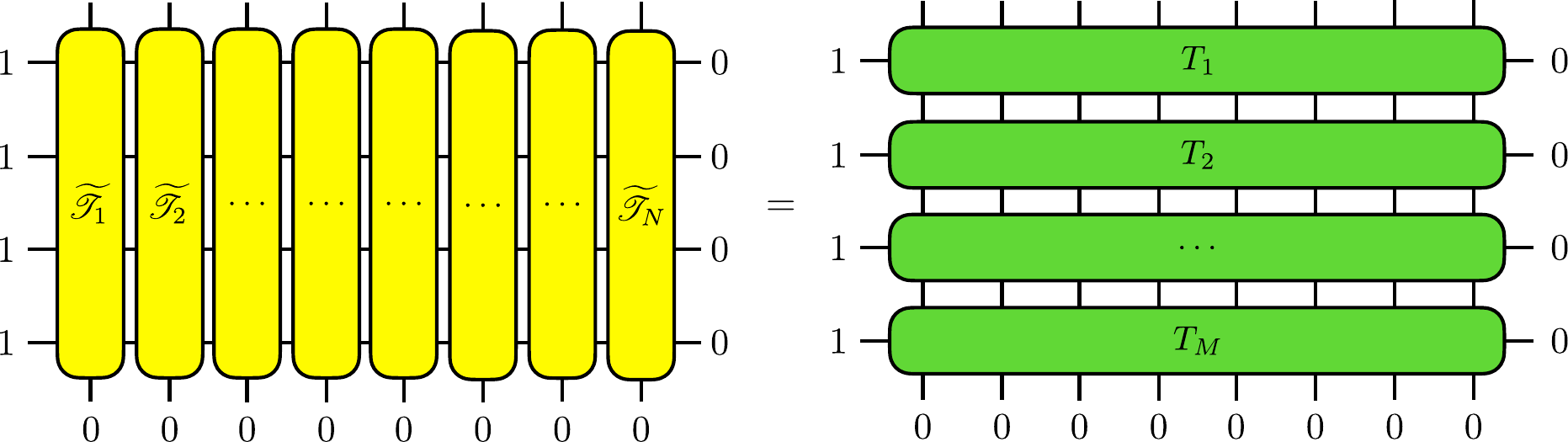}
    \caption{Bethe state according to
    the ABA. The standard formulation of the ABA (\ref{ABAMNdir}) based on $T_a$
    equals the dual formulation of the ABA
    in the F-basis (\ref{ABAMNdual})
    based on $\widetilde{\mathscr{T}}_j$.}
    \label{figABAMN}
\end{figure}

We begin with
the unnormalized Bethe state
of $M$ magnons over $N$ spins
as per the ABA~\cite{Korepin93}:
\begin{equation}
    \label{ABAMNdir}
    B(u_1)\ldots B(u_M)\ket{0}^{\otimes N}=
    \prod_{a = 1}^M\left[\vphantom{\prod_{j = 1}^N}\!\bra{0}_aT_a^{\vphantom{x}}\ket{1}_{a}\!\right] \ket{0}^{\otimes N} =
    \prod_{a = 1}^M\left[\bra{0}_a \prod_{j = 0}^{N-1} R_{aN-j}\ket{1}_a\right] \ket{0}^{\otimes N}\ ,
\end{equation}
where $B(u_a)$ are the operators in (\ref{ABA}),
which commute among themselves,
$\ket{i}_a$ belongs to the Hilbert space of the $a$-th ancilla,
and $\ket{0}^{\otimes N}$ is the reference state in the quantum space.
The so-called ``magnons'' are 
to be identified with plane waves, as
the CBA
makes clear.
(See (\ref{MPS}) below.)
If we rearrange the product of R-matrices, 
according to (\ref{FMTdual}) and (\ref{symTdual}),
we can write 
\begin{equation}
    \label{ABAMNdual}
    B(u_1)\ldots B(u_M)\ket{0}^{\otimes N}
    = \sum_{i_j=0,1}
    \bra{0}^{\otimes M}\!
    \widetilde{\mathscr{T}}_{N}^{\,i_N}\ldots\widetilde{\mathscr{T}}_{2}^{\,i_2}\widetilde{\mathscr{T}}_{1}^{\,i_1}
    \ket{1}^{\otimes M}
    \ket{i_1 \ldots i_N}  \ ,
\end{equation}
where $\ket{i}_j$ belongs to the Hilbert space of the $j$-th spin, $\ket{i}^{\otimes M}$ belongs to the auxiliary space, we introduced the following notation for the first column of (\ref{ABAdual}):
\begin{equation}
    \widetilde{\mathscr{T}}_{j}^{\,0}:=  \widetilde{\mathscr{A}}_{j} \ , \quad   \widetilde{\mathscr{T}}_{j}^{\,1}= \widetilde{\mathscr{C}}_{j} \ ;
\end{equation}
and we used
\begin{equation}
    \label{Fonbs}
    F_{12\ldots M}\ket{i}^{\otimes M}=\ket{i}^{\otimes M}\ .
\end{equation}
Figure \ref{figABAMN} depicts the Bethe state according to the ABA. 
It is worth noting that Figure \ref{figABAMN}, corresponding to (\ref{ABAMNdual}), can be graphically obtained   
from Figures~\ref{figMT}--\ref{figsymTdual}  
by taking into account the relation~(\ref{Fonbs}).
The Bethe state (\ref{ABAMNdual}) is an MPS with tensors $\widetilde{\mathscr{T}}_{j}^{\,i}$ and bond dimension $2^M$.

MPSs represent a highly structured form to describe many-body states in one-dimension~\cite{Cirac20}. 
MPSs, in particular, realize states as sequences of matrices defined over an auxiliary space, 
each associated with a spin, 
such that the global wave function emerges as a product of these spin-specific matrices. 
The dimension of the matrices connecting adjacent spins, called ``tensors'', is known as the ``bond dimension''. 
The bond dimension quantifies the entanglement between bipartitions of the state across the boundary between spins.
The MPS representation of a given state is inherently non-unique. 
The matrices at individual spins can be modified by a gauge transformation in the auxiliary space, 
which leaves the overall wave function unchanged.
For instance, 
let $V_j$ be
the invertible
$2^M\times 2^M$-matrices 
of a gauge transformation acting on the auxiliary space of (\ref{ABAMNdual}).
The mapping
\begin{equation}
    \widetilde{\mathscr{T}}_{j}^{\,i}\mapsto V_{j}^{-1}\widetilde{\mathscr{T}}_{j}^{\,i}\,V_{j-1} \ , \quad \bra{0}^{\otimes M}\mapsto\bra{0}^{\otimes M} V_{N} \ , \quad  \ket{1}^{\otimes M}\mapsto V_{0}^{-1}\ket{1}^{\otimes M} \ ,
\end{equation}
yields another admissible MPS representation of the Bethe state (\ref{ABAMNdual}).
If all the matrices of the gauge transformation are equal,
the transformation is called ``global'',
otherwise it is called ``local''.

The discussion up to this point applies to both the homogeneous and inhomogeneous XXZ models.
To align with \cite{Ruiz23},
whose spin chain is homogeneous,
we set $v_j=0$, hence
$\widetilde{\mathscr{T}}_{j}^{\,i}=\widetilde{\mathscr{T}}^{\,i}$
until the end of the subsection.
The MPS of the Bethe state becomes uniform,
which means that all the tensors of (\ref{ABAMNdual}) are equal.

Let us derive the representation of the CBA as an MPS from (\ref{ABAMNdual}).
We perform a global gauge transformation by the $2^M\times 2^M$-matrix
\begin{equation}
    \label{V}
    V={\overset{M}{\underset{a=1}{\bigotimes}}}\begin{bmatrix}
        g_a & 0 \\
        0 &  \prod_{b=1,\ b\neq a}^M f_{ab} 
    \end{bmatrix}
    \ ,
\end{equation}
where 
\begin{equation}
    f_a:=f(u_a) \ , \quad g_a:=g(u_a) \ .
\end{equation}
(We introduced $f_a$ already for convenience.)
The global gauge transformation just rescales
$\widetilde{\mathscr{T}}^{\,1}$ 
because $\widetilde{\mathscr{T}}^{\,0}$ is diagonal.
The tensors of the MPS representation become
\begin{equation}
    \label{Lambda}
    \begin{split}
        \Lambda^0&= V^{-1}\widetilde{\mathscr{T}}^{\,0}V
        =\overset{M}{\underset{a=1}{\bigotimes}}
        \begin{bmatrix}
        1 & 0 \\
        0 & f_a
        \end{bmatrix} \ , \\
        \Lambda^1&= V^{-1}\widetilde{\mathscr{T}}^{\,1}V=
    \sum_{a=1}^{M}
    \overset{a-1}{\underset{b=1}{\bigotimes}}
    \begin{bmatrix}
        1 & 0 \\
        0 & f_{ab}f_b 
    \end{bmatrix}
    \begin{bmatrix}
        0 & 1 \\
        0 & 0
    \end{bmatrix}
    \overset{M}{\underset{c=a+1}{\bigotimes}}
    \begin{bmatrix}
        1 & 0 \\
        0 & f_{ac}f_c 
    \end{bmatrix}
    \ .
    \end{split}
\end{equation}
The transformation
leaves a multiplicative factor due to the action of $V^{-1}$ and $V$ on
$|0\rangle^{\otimes M}$ and~$|1\rangle^{\otimes M}$, respectively.
If we re-define
\begin{equation}
    |\Psi^{[M]}_N\rangle=\left[ \prod_{a=1}^M\,\frac{
    1
    }{ g_a }\,\underset{b\neq a}{\prod_{b=1}^M}\,f_{ab}\,
    \right] B(u_1)\ldots B(u_M)\ket{0}^{\otimes N} \ ,
\end{equation}
we obtain
\begin{align}
    \label{MPS}
        |\Psi_{N}^{[M]} \rangle&=
        \sum_{i_j=0,1}
        \bra{0}^{\otimes M}\Lambda^{i_N}\ldots\Lambda^{i_2}\Lambda^{i_1}\ket{1}^{\otimes M}
        \ket{i_1 \ldots i_N} \\
        &=\underset{1\leq n_1<\ldots<n_M\leq N}{\sum}
        \underset{a_p\neq a_q}{\sum_{a_1,\ldots,a_M=1}^M}
       \left[\underset{1\leq q<p\leq M}{\prod}s_{a_q a_p }\right]
        \left[\overset{M}{\underset{p=1}{\prod}} \,\, x_{a_p}^{n_p -1}\right]
        | n_1 \ldots n_M \rangle_N \ , \nonumber
\end{align}
where
\begin{equation}
    |n_1 \ldots n_M\rangle_N=\sigma_{n_1}^{-}\ldots\sigma_{n_M}^{-} 
        \ket{0}^{\otimes N} \ .
\end{equation}
The diagonal elements of the R-matrix (\ref{Rmatrix}) 
define the quasi-momenta $p_1,\ldots, p_M$ of the magnons:
\begin{equation}
    \label{qmomenta}
    x_a=\exp(\im p_a):=f_a=\frac{\sinh u_a}{\sinh(u_a+\im\gamma)} \ .
\end{equation}
The scattering amplitudes are
\begin{equation}
    \label{scattering}
    s_{ab}:=f_{ab}=\frac{\sinh(u_a-u_b)}{\sinh(u_a-u_b+\im\gamma)} \ ,
\end{equation}
in terms of which the two-body S-matrix reads
\begin{equation}
    \label{Smatrix}
    S_{ab}=\frac{s_{ba}}{s_{ab}}=-\frac{\sinh(u_a-u_b+\im\gamma)}{\sinh(u_b-u_a+\im\gamma)} \ .
\end{equation}
Since the Bethe state (\ref{MPS}) realizes the unnormalized linear superposition of $M$
magnons with quasi-momenta $p_1,p_2\ldots, p_M$ over $N$ spins,
we call it the ``MPS
of the CBA''. 
The proof of the equivalence between the first and second lines of (\ref{MPS}) appears in Appendix~B of~\cite{Ruiz23}.  
(We provide the general proof for the Bethe states of the inhomogeneous spin chain  
in Appendix~\ref{appMPSCBA}.)
We stress that neither the MPS of the ABA~(\ref{ABAMNdual})  
nor its reformulation as an MPS of the CBA~(\ref{MPS})  
requires $u_a$ to satisfy the Bethe equations.  
Therefore, the Bethe states under consideration are not necessarily eigenstates  
of the Hamiltonian of the homogeneous XXZ model~(\ref{H}).

Reference 
\cite{Ruiz23} 
used an equivalent MPS of the CBA. 
We prove the equivalence between both representations in Appendix \ref{appMPSCBA}.
By changing its initialization in the auxiliary space, the MPS of \cite{Ruiz23} 
realizes Bethe states with $0\leq r\leq M$ magnons 
and support on $1\leq k\leq N$ qubits.
This fact underpinned the 
construction of the
unitaries
of ABCs in \cite{Ruiz23}.
The MPS
(\ref{MPS}) 
also permits to define these Bethe states in the same fashion.
A Bethe state of 
$r$ 
magnons over $k$ spins with quasi-momenta
$p_{m_a}$ 
chosen
among 
$p_a$
is
\begin{align}
    \label{partial}
    |\Psi_{k}^{[r]} \rangle  &=
    \sum_{i_j=0,1}
    \bra{0}^{\otimes M}
    \,
    \Lambda^{i_k}\ldots\Lambda^{i_2}\Lambda^{i_1}
    \ket{m_1\ldots m_r}_{M}
    \ket{i_1 \ldots i_k} \\
    &=\underset{1\leq n_1<\ldots<n_r\leq k}{\sum}\,
    \underset{a_p\neq a_q}{\sum_{a_1,\ldots,a_r=1}^r}
   \left[\underset{1\leq q<p\leq M}{\prod}s_{m_{a_q} m_{a_p} }\right]
    \left[
    \overset{M}{\underset{p=1}{\prod}} \,\, x_{m_{a_p}}^{n_p -1}\right]
    | n_1 \ldots n_r \rangle_k \ . \nonumber
\end{align}
Equation (\ref{partial}) states that the MPS of the CBA 
initialized on 
$|m_1 \ldots m_r\rangle_M$ 
in the auxiliary space, for every number of 
tensors~$k$, 
realizes a Bethe state whose quasi-momenta are determined by this initial state. 
The correspondence between Bethe states and elements of the computational basis of the auxiliary space is one-to-one for 
fixed~$k$. 
We also note that 
(\ref{partial})
does not hold for the MPS representation of the ABA in terms of
$\mathscr{T}^i$.
The demonstration of (\ref{partial})
follows from a direct computation, 
but there is an alternative derivation of this property.
We end this subsection by proving that
(\ref{partial})
is a consequence
of the symmetry with respect to
exchange of
ancillae 
of dual monodromy matrices 
in the F-basis (\ref{symTdualprop}).
The proof also highlights the convenience of the change of normalization 
(\ref{Lambda}),
as different Bethe states would carry different normalizations if they were defined by the MPS of
$\widetilde{\mathscr{T}}^{\,i}$.

\begin{figure}[ht]
        \centering
        \includegraphics[width=\textwidth]{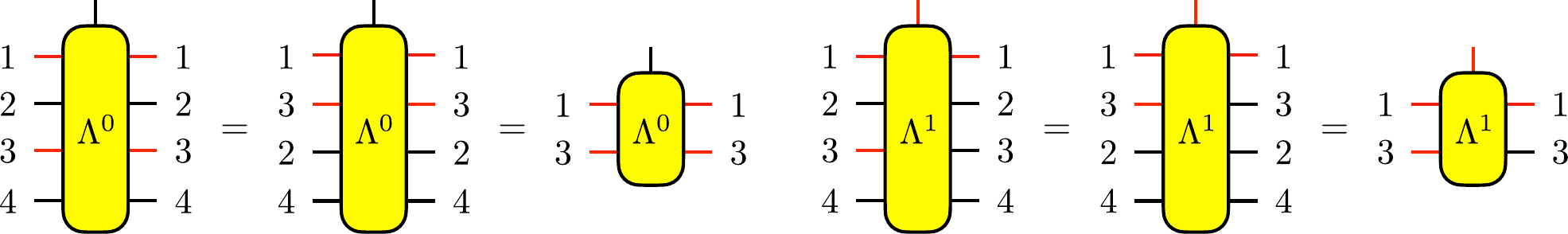}
        \caption{
            Application of the symmetry with respect to exchange of ancillae to
            $\Lambda^i$.
            Black and red lines represent
            qubits on $\ket{0}$ and
            $\ket{1}$, respectively.
            Vertical and horizontal lines correspond to spins and ancillae, respectively.
            Numbers besides
            horizontal lines denote the $u_a$
            identifying the ancillae.
        }
        \label{figLambda}
\end{figure}

Consider (\ref{symTdualprop}).
The gauge-transformation
matrix (\ref{V}) is diagonal.
The effect
of the gauge
transformation on
the matrix elements 
of $\widetilde{\mathscr{T}}^{\,i}$
is just a rescaling,
as we already mentioned.
The rescaling does not spoil 
the symmetry with respect to the
exchange of
ancillae, which
still holds for (\ref{Lambda}). Therefore,
\begin{equation}
    \begin{split}
        \label{Lambdahomprop}
        \bra{i_1 \ldots i_M}\Lambda^i(u_1,\ldots,u_M)&\ket{j_1\ldots j_M}
       =\bra{i_{\tau_1}\ldots i_{\tau_M}}\Lambda^i(u_{\tau_1},\ldots,u_{\tau_M})
        \ket{j_{\tau_1}\ldots j_{\tau_M}} \ ,
    \end{split}
\end{equation}
for every permutation $\tau\in S_M$.
It is clear from (\ref{Lambda}) that
$\Lambda^{i}$ only have
non-trivial matrix elements
between states of
the form $|m_1 \ldots m_r\rangle_M$ and $|n_1 \ldots n_{r-i}\rangle_M$.
Furthermore,
the contribution
to the matrix elements between 
ancillae 
on $\ket{0}$ is a multiplicative
factor of one.
This feature, 
which differentiates 
$\Lambda^i$
from
$\widetilde{\mathscr{T}}^{\,i}$,
enables the reduction of the number of ancilla in the auxiliary space. 
We can write the action of the first $\Lambda^i$ on the initial state of (\ref{partial}) like
\begin{equation}
    \label{Lambda0prop}
    \begin{split}
    \Lambda^{0}(u_1,\ldots,u_M)|m_1 \ldots m_r\rangle_M
    =\,\langle 1|^{\otimes r} \, 
    \Lambda^{0}(u_{m_1},\ldots,u_{m_r})|1\rangle^{\otimes r}  
    |m_1 \ldots m_{r}\rangle_M \ , \\
    \end{split}
\end{equation}
and
\begin{equation}
    \label{Lambda1prop}
    \begin{split}
    \Lambda^{1}(u_1,\ldots,u_M)|m_1 \ldots m_r\rangle_M
    =\sum_{a=1}^r\,\langle 1|^{\otimes r}
    \sigma_a^- \, 
    \Lambda^{1}(u_{m_1},\ldots,u_{m_r})|1\rangle^{\otimes r}  
    \sigma_{m_a}^+|m_1 \ldots m_r\rangle_M \ , 
    \end{split}
\end{equation}
where we used
(\ref{Lambdahomprop})
with a permutation $\tau$ that verifies
\begin{equation}
    \begin{split}
    \tau_{a}=m_a \ , \quad a=1,2,\ldots,r \ ,
    \end{split}
\end{equation}
but is otherwise arbitrary.
We depict 
(\ref{Lambda0prop}) 
and 
(\ref{Lambda1prop}) in 
Figure 
\ref{figLambda}.
The concatenation of
(\ref{Lambda0prop}) and (\ref{Lambda1prop}) in
the Bethe wave function of (\ref{partial}) leads us to
\begin{equation}
    \label{hompartial}
    \begin{split}
    \bra{0}^{\otimes M}
    \Lambda^{i_k}(u_1,\ldots,u_M)&\ldots
    \Lambda^{i_1}(u_1,\ldots,u_M)
    |m_1 \ldots m_r\rangle_M
    \\
    =&\bra{0}^{\otimes r}
    \!\,
    \Lambda^{i_k}(u_{m_1},\ldots,u_{m_r})\ldots
    \Lambda^{i_1}(u_{m_1},\ldots,u_{m_r})
    \ket{1}^{\otimes r} \ .
    \end{split}
\end{equation}
Therefore,
if we reorder the quasi-momenta in $\Lambda^i$
and eliminate the ancillae that remain on $|0\rangle$ out,
we can prove that Bethe states 
with a few magnons have the form
(\ref{MPS}).
Note that the proof is a consequence 
of the definitorial property of the F-basis
(\ref{symTdualprop}).

Given an MPS with the property 
(\ref{partial}),
one can construct the unitaries of ABCs in the scheme of 
\cite{Ruiz23}.
Motivated by the connection between the CBA and the F-basis we uncovered,
we now turn to the inhomogeneous spin chain to construct their ABCs.

\subsection{The F-basis and Coordinate Bethe Ansatz: Inhomogeneous Spin Chain}

\label{ssCBAinh}

In this subsection, 
we compute the MPS formulation of the CBA for the inhomogeneous XXZ model.
This representation leads to the exact unitaries preparing normalized Bethe states in Subsection \ref{ssABCinh}.

Let us turn back to
(\ref{ABAMNdual}).
The formula says Bethe states of the ABA are MPSs whose tensors are $\widetilde{\mathscr{T}}^{i}_j$.
In Subsection
\ref{ssCBAhom},
we performed a global gauge transformation to obtain the tensor~$\Lambda^i$
for~\mbox{$v_j=0$}, 
when the MPS is uniform and the spin chain homogeneous. 
To obtain analogous tensors for general $v_j$,
we promote~(\ref{V}) 
to the $2^M\times 2^M$-matrices of a local gauge transformation:
\begin{equation}
    \label{Vj}    
    V_j
    ={\overset{M}{\underset{a=1}{\bigotimes}}}
    \begin{bmatrix}
        g_{aj} & 0 \\
        0 &  \prod_{b=1,\ b\neq a}^M f_{ab} 
    \end{bmatrix}
    \ .
\end{equation}
The transformation acts on the tensors $\widetilde{\mathscr{T}}^{i}_j$ like
\begin{equation}
    \label{Lambdaj01}
    \begin{split}
        \Lambda^0_j&= V_j^{-1}\widetilde{\mathscr{T}}_j^{\,0}V_{j-1}
        =\overset{M}{\underset{a=1}{\bigotimes}}
        \begin{bmatrix}
        1 & 0 \\
        0 & f_{aj}
        \end{bmatrix} \ , \\
        \Lambda^1_j&= V_j^{-1}\widetilde{\mathscr{T}}_j^{\,1}V_{j-1}=
    \sum_{a=1}^{M}
    \overset{a-1}{\underset{b=1}{\bigotimes}}
    \begin{bmatrix}
        1 & 0 \\
        0 & f_{ab}f_{bj}
    \end{bmatrix}
    \begin{bmatrix}
        0 & 1 \\
        0 & 0
    \end{bmatrix}
    \overset{M}{\underset{c=a+1}{\bigotimes}}
    \begin{bmatrix}
        1 & 0 \\
        0 & f_{ac}f_{cj}
    \end{bmatrix}
    \ .
    \end{split}
\end{equation}
The action of
$V_0$ 
on 
$|1\rangle^{\otimes M}$
and 
$V_{N-1}^{-1}$ 
on 
$|0\rangle^{\otimes M}$ 
produces an overall normalization that we cancel by redefining the Bethe state like
\begin{equation}
    |\Psi^{[M]}_N\rangle=\left[ \prod_{a=1}^M\,\frac{
    1
    }{ g_{aN-1} }\,\underset{b\neq a}{\prod_{b=1}^M}\,f_{ab}\,
    \right] B(u_1)\ldots B(u_M)\ket{0}^{\otimes N}\ .
\end{equation}
Explicitly,
\begin{align}
    \label{inhMPS}
    |\Psi_{N}^{[M]} \rangle&=
        \sum_{i_j=0,1}
        \bra{0}^{\otimes M}\Lambda^{i_N}_N\ldots \Lambda^{i_2}_2\Lambda^{i_1}_1\ket{1}^{\otimes M}
        \ket{i_1 \ldots i_N} \\
        &=\underset{1\leq n_1<\ldots<n_M\leq N}{\sum}\
        \underset{a_p\neq a_q}{\sum_{a_1,\ldots,a_M=1}^M}
        \left[\underset{1\leq q<p\leq M}{\prod}s_{a_q a_p }\right]
        \left[\vphantom{\underset{1\leq a<b\leq M}{\prod}}
        \overset{M}{\underset{p=1}{\prod}} \overset{n_p-1}{\underset{j=1}{\prod}} \,\, x_{a_p,j}\right]
        | n_1\ldots n_M \rangle_N \ . \nonumber
\end{align}
The lack of uniformity of the MPS above is reflected in the emergence of position-dependent quasi-momenta:
\begin{equation}
    \label{xaj}
    x_{a,j}=
    \exp(\im p_{a,j}):=
    f_{aj}=
    \frac{\sinh(u_a-v_j)}{\sinh(u_a-v_j+\im\gamma)} \ .
\end{equation}
We assume that $u_a$, 
and consequently $p_{a,j}$, 
do not necessarily satisfy the Bethe equations. 
This means that the Bethe state (\ref{inhMPS}) is not in general an eigenstate of the transfer matrix that defines the inhomogeneous spin chain.
We also assume that 
$u_a\neq u_b$,
ensuring that quasi-momenta are distinct at each position so that the Bethe wave function does not vanish identically.
The scattering amplitudes are unchanged with respect to the homogeneous spin chain and 
equal~(\ref{scattering}). 
Since
(\ref{inhMPS})
realizes a Bethe wave function 
with~$M$ 
magnons propagating 
over~$N$ 
spins in the inhomogeneous spin chain,
we identify it with the ``MPS of the CBA''.
Magnons are still plane waves whose scattering is governed by the two-body S-matrix
(\ref{Smatrix}),
but their quasi-momenta vary as the plane wave propagates through spin chain.
The parameterization~(\ref{inhMPS})
directly generalizes the CBA for the homogeneous spin chain~(\ref{MPS}).
For instance, if $M=2$:
\begin{equation}
    |\Psi^{[2]}_N\rangle
    =\underset{1\leq n_1<n_2\leq N}{\sum}
    \left(
    s_{12}
    \left[\overset{n_1-1}{\underset{j=1}{\prod}} x_{1,j}\right]
    \left[\overset{n_2-1}{\underset{k=1}{\prod}} x_{2,k}\right]
    +s_{21}
    \left[\overset{n_1-1}{\underset{j=1}{\prod}} x_{2,j}\right]
    \left[\overset{n_2-1}{\underset{k=1}{\prod}} x_{1,k}\right]
    \right)|n_1n_2\rangle \ .
\end{equation}
The proof of the equivalence between the first and second line of (\ref{inhMPS}) appears in 
Appendix~\ref{appMPSCBA}. 

Like
(\ref{MPS}), 
the MPS of the CBA
(\ref{inhMPS})
enables the systematic construction of Bethe with less than $M$ magnons~\footnote{
    The MPS in terms of $\widetilde{\mathscr{T}}_j^{i}$~(\ref{ABAMNdual}) also gives rise to Bethe states with a few magnons thanks to the symmetry~(\ref{symTdualprop}).
    However,
    normalizations are more intricate,
    which complicates the derivation of unitaries.
    }.
This fact is the upshot of 
(\ref{symTdualprop}), 
the symmetry of dual monodromy matrices with respect to the exchange of ancillae in the F-basis. 
Following the steps of
Subsection~\ref{ssCBAhom}, 
we deduce 
\begin{equation}
    \label{inhompartial}
    \begin{split}
    \bra{0}^{\otimes M}
    \Lambda^{i_k}_k(v_k;u_1,\ldots,u_M)&\ldots
    \Lambda^{i_1}_1(v_1;u_1,\ldots,u_M)
    |m_1 \ldots m_r\rangle_M
    \\
    =&\bra{0}^{\otimes r}
    \!\,
    \Lambda^{i_k}_k(v_k;u_{m_1},\ldots,u_{m_r})\ldots
    \Lambda^{i_1}_1(v_1;u_{m_1},\ldots,u_{m_r})
    \ket{1}^{\otimes r} \ .
    \end{split}
\end{equation}
We emphasize that the importance of (\ref{inhompartial}), alongside (\ref{hompartial}),  
lies in the fact that, by modifying the initialization of the MPS in the auxiliary space,  
the MPS also realizes Bethe states with~\mbox{$0 \leq r \leq M$} magnons and support on $1 \leq k \leq N$ qubits.  
This property is closely related to the equality between the MPS  
and the linear superpositions of plane waves of the CBA in (\ref{inhMPS}).  

We should mention 
\cite{Ovchinnikov10}
already stressed the suitability of the F-basis to compute Bethe wave functions.
We clarify the relation between
(\ref{inhMPS})
and the parameterizations of the Bethe wave functions for the inhomogeneous spin chain of
\cite{Ovchinnikov10} 
in Appendix \ref{appMPSCBA}.

\subsection{Inhomogeneous Algebraic Bethe Circuits}

\label{ssABCinh}

Having obtained the MPS of the CBA 
(\ref{inhMPS}) with the central property 
(\ref{inhompartial}), 
we are in position to construct ABCs for the inhomogeneous spin chain.
The first step is the decomposition of the Hilbert space of 
$k$ 
qubits into eigenspaces of definite total spin along the 
$z$-axis:
\begin{equation}
    \label{orthsum}
    \mathsf{H}_{k}= \overset{k}{\underset{r=0}{\bigoplus}}~\mathsf{H}_k^{[r]} \ ,
\end{equation}
with
\begin{equation}
    \begin{split}
    \label{Hk}
        \mathsf{H}_k^{[r]}&= 
        \mathrm{span}{\left\{\ket{i_1\ldots i_k} \, | \, 
        \sum_{j=1}^k i_j = r \right\}} =\mathrm{span}{\left\{
            \vphantom{\sum_{j=1}^M}
        \ket{n_1 \ldots n_r}_k \, |
         \, 1\leq n_1<\ldots<n_r\leq k\right\}} \ .
    \end{split}
\end{equation}
The dimension of the eigenspace is
\begin{equation}
    \label{dimHA}
        \dim \mathsf{H}_k^{[r]} = \binom{k}{r} \ ,
\end{equation}
where we define the binomial coefficient to vanish if $r<0$ or $r>k$.
We stress
(\ref{Hk})
encompasses as particular cases the quantum
(\ref{HN})
and auxiliary
(\ref{HM})
Hilbert spaces for
$k=N$
and
$k=M$, 
respectively.
We introduced the decomposition because Bethe states arrange into eigenspaces of definite total spin along the 
$z$-axis,
thus the unitaries of ABCs based on them break into blocks.
Bethe states 
with~$r$
magnons over 
$k$ 
spins specifically belong to
$\mathsf{H}_k^{[r]}$.
If 
$k$ 
and 
$r$
with~$r\leq k$
are fixed,
Bethe states are in one-to-one correspondence with elements of the computational basis of~$\mathsf{H}_k^{[r]}$.
In 
Appendix~\ref{appindexation},
we define the indices
$\alpha$ 
to label strings
$1\leq n_1<\ldots<n_r\leq k$ 
inside the eigenspace~$\mathsf{H}_k^{[r]}$.
We use the index to label the objects of ABCs.
We do not 
allow~$r$ 
to be greater than 
$M$,
although~$M$
itself could belong to
$0\leq M\leq N$.

\begin{figure}[ht]
    \centering
    \includegraphics[width=0.675\textwidth]{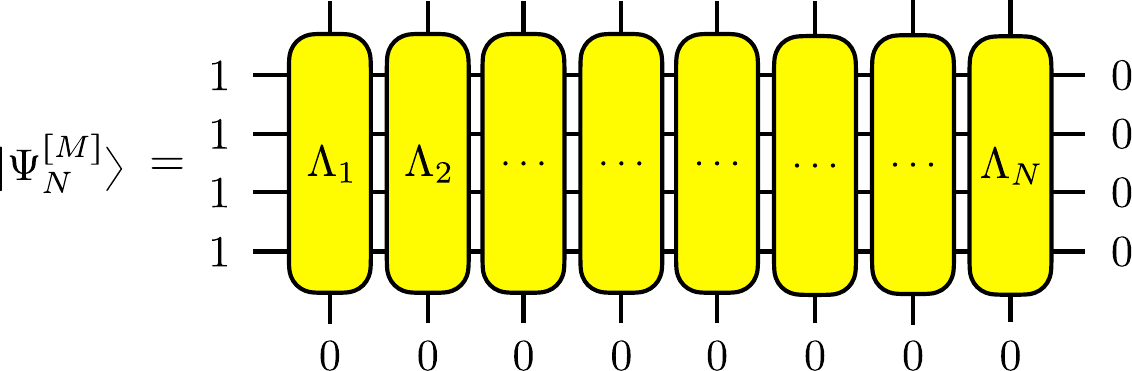}
    \caption{Bethe state as the non-uniform MPS of the CBA in the inhomogeneous spin chain.
    The Bethe state is a linear superposition of plane waves with spin-dependent quasi-momenta.
    }
    \label{figPsiMNinh}
\end{figure}

The arrangement of Bethe states into eigenspaces of definite total spin along the $z$-axis is clarified by the MPS of the CBA.
To see this,
it is convenient to assemble both $\Lambda^0$ and $\Lambda^1$ into the non-unitary tensor
\begin{equation}
    \label{Lambdaj}
    \Lambda_j=\Lambda_j(v_j;u_1,\ldots,u_M):
    \mathsf{H}_M\otimes\mathsf{h}_j\cong \mathsf{H}_{M+1}
    \rightarrow
    \mathsf{H}_{M+1}\cong\mathsf{h}_j\otimes\mathsf{H}_M \ , \quad \Lambda_j^i:=\bra{i}_j\Lambda_j\ket{0}_j \ ,
\end{equation}
where $|i\rangle_j$ belongs to the Hilbert space of the $j$-th spin.
We have defined~$\Lambda_j$ 
so that it moves the position of~$\mathsf{h}_j$
in the tensor product from the last to the first place, 
which is convenient for the derivation of ABCs.
Note we have not specified~$\Lambda_j|1\rangle_j$ 
as it plays no role in the MPS.
The 
tensor~$\Lambda_j$ 
by construction commutes
and the total spin along 
$z$-axis of 
$M+1$ 
qubits
($M$ 
ancillae and the 
$j$-th spin 
of the quantum space). 
Since the product of tensors 
in the auxiliary space of the MPS~(\ref{inhMPS}) 
is initialized 
on~$\ket{1}^{\otimes M}$ 
and projected 
on~$\bra{0}^{\otimes M}$, 
the Bethe state must carry a definite number of ones.
We depict the MPS~(\ref{inhMPS}) in terms of~$\Lambda_j$
in Figure~\ref{figPsiMNinh}.

Therefore, we can express
\begin{equation}
    \label{blocksLambda}
    \begin{split}
        \Lambda^i_j&= \overset{M}{\underset{r=0}{\bigoplus}}~\Lambda^{[i,r]}_j \ , \quad  \Lambda^{[i,r]}:\mathsf{H}_M^{[r]}\rightarrow \mathsf{H}_M^{[r-i]} \ .
    \end{split}
\end{equation}
The number of rows and columns of the non-unitary matrices is
\begin{equation}
    \#~\mathrm{rows} \times \#~\mathrm{columns~of}~\Lambda_{j}^{[i,r]} =\binom{M}{r-i}  \times \binom{M}{r}  \ .
\end{equation}
The square block 
$\Lambda^{[0,r]}$
is diagonal with entries
\begin{equation}
    \label{Lambda0r}
    \Lambda_{j,\alpha\beta}^{[0,r]}=\delta_{\alpha\beta}
    \overset{r}{\underset{p=1}{\prod}} \,x_{n_p,j}\ ,
\end{equation}
where 
\begin{equation}
    \label{alphaequivm}
    |\alpha\rangle=|m_1\ldots m_r\rangle_k \ , \quad 
    |\beta\rangle=|n_1\ldots n_r\rangle_k \ ,
\end{equation}
with $k=M$.
The rectangular block 
$\Lambda^{[1,r]}$ 
has the entries
\begin{equation}
    \label{Lambda1r}
    \Lambda^{[1,r]}_{j,\alpha_r\beta}=\sum_{p=1}^{r}\delta_{\alpha_r\beta_p} \underset{q\neq p}{\overset{r}{\underset{q=1}{\prod}}}\, s_{n_p n_q}x_{n_q,j} \ ,
\end{equation}
where we identified 
\begin{equation}
    \label{alphar}
    |\alpha_r\rangle=|m_1\ldots \ldots m_{r-1}\rangle_{k-1} \ , \quad |\beta_p\rangle=|n_1\ldots n_{p-1}n_{p+1}\ldots n_r\rangle_k \ ,
\end{equation}
with $k=M$.
Given the block decomposition of 
$\Lambda_j$, 
we seek to write the MPS of the CBA
(\ref{inhMPS})
as a quantum circuit.
In other words,
we want to compute unitaries out of the non-unitary tensors 
$(\ref{Lambdaj})$. 
The unitaries themselves decompose into unitary blocks of definite total spin along the $z$-axis.  
We achieve this goal by a local gauge transformation that puts 
(\ref{inhMPS})
into the canonical form.
The canonical form is a standard representation of an MPS where the tensor is subject to orthonormalization constraints~\cite{Cirac20}.
The tensor in the canonical form is unique up to unitary rotations in the auxiliary space.
Here we use the left canonical form of the MPS (\ref{inhMPS}),
defined as that whose 
tensors~$\Lambda_j\ket{0}_j$ 
become isometries with more rows than columns. 

\begin{figure}[ht]
    \centering
    \includegraphics[width=0.7\textwidth]{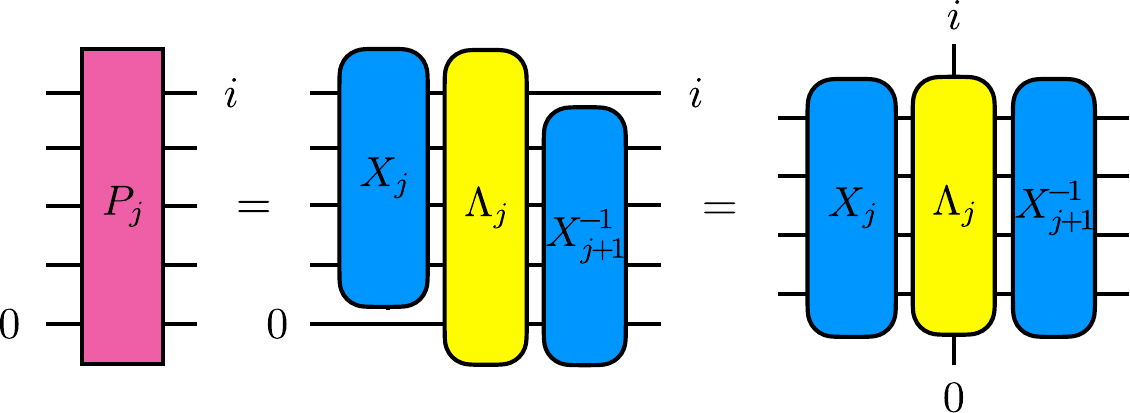}
    \caption{
    $M+1$-qubit (sharp-cornered)
    unitaries
    from the (rounded-cornered) non-unitary tensor 
    by a gauge transformation as per (\ref{Pjgauge}).
    Locating ancillae and spins in a single array is necessary to obtain ABCs, 
    whose unitaries follow from the elimination of post-selected qubits.
    }
    \label{figPLambda}
\end{figure}

The set of matrices
$X_j$
realize the local gauge transformation on the auxiliary space we want.
The 
$M+1$-qubit 
unitary is built as
\begin{equation}
    \label{Pjgauge}
    P_j= X_{j+1}^{-1} \Lambda_j X_j \ .
\end{equation}
We depict (\ref{Pjgauge}) in Figure \ref{figPLambda}. Unitarity holds if
\begin{equation}
    \label{unitarityP}
    P_j^\dagger P_j = 1_{2^{M+1}} \ .
\end{equation}
We assume 
$X_j$ 
preserve the total spin along the 
$z$-axis, 
\begin{equation}
    X_j=\overset{M+1}{\underset{r=0}{\bigoplus}}\,X_j^{[r]} \ , 
\end{equation}
hence
$P_j$ 
decomposes into unitary blocks according to 
$r$. (See (\ref{longdirectsum}) and (\ref{shortdirectsum}) below.)

\begin{figure}[ht]
    \centering
    \includegraphics[width=0.8\textwidth]{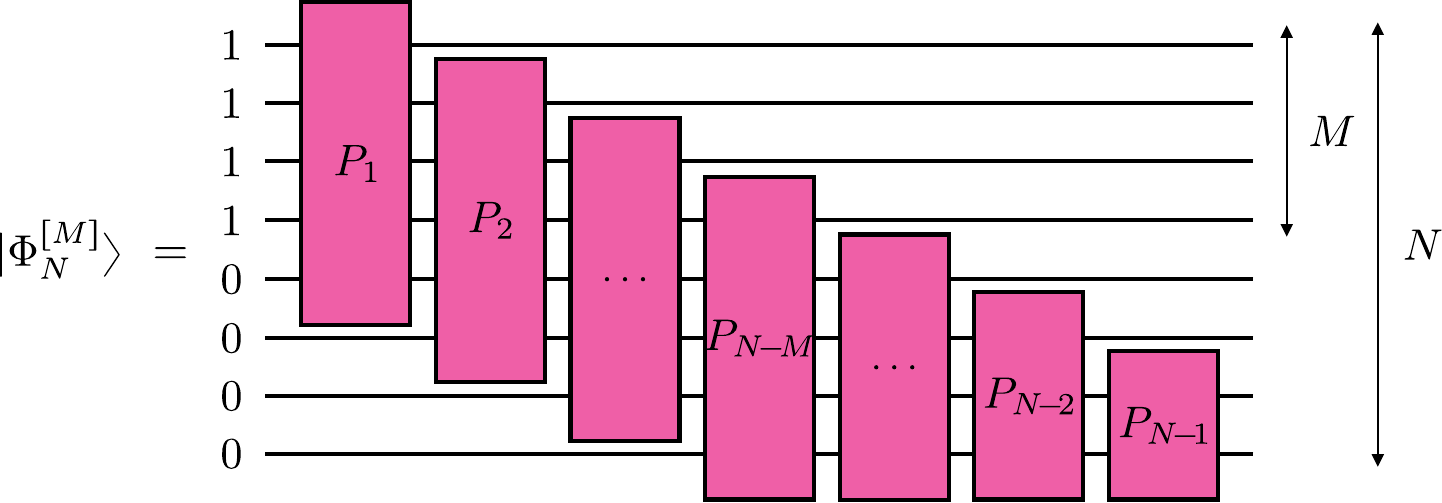}
    \caption{
    ABCs are deterministic algorithms preparing normalized Bethe states of the inhomogeneous XXZ model.
    The unitaries are either long or short.
    }
    \label{figABC}
\end{figure}

The local gauge transformation 
(\ref{Pjgauge}) 
leads us to a quantum circuit.
However, 
the quantum algorithm thus obtained is probabilistic.
The last $M$ qubits of the $N+M$ qubits must be \mbox{post-selected} on 
$\ket{0}^{\otimes M}$.
To avoid the computational cost of post-selection,
we eliminate these qubits.
The result is the circuit of ABCs,
a class of deterministic quantum algorithms.
We still denote the unitaries by 
$P_j$,
despite that the elimination of post-selected qubits reduces the size of the last $M$ unitaries in the circuit.
ABCs prepare normalized Bethe states of~$M$ magnons over~$N$ spins:
\begin{equation}
    |\Phi^{[M]}_N\rangle=
    \frac{1}{ \langle\Psi^{[M]}_N|\Psi^{[M]}_N\rangle} 
    |\Psi^{[M]}_N\rangle=
    P_{N-1}\ldots P_2 P_1
    \ket{1}^{\otimes M}\ket{0}^{\otimes N-M} \ .
\end{equation}
We depict the circuit in 
Figure \ref{figABC}.

The number of qubits on which the unitaries act defines two classes:
long 
($1\leq j\leq N-M$) 
and short 
($N-M< j\leq N$) 
unitaries. 
The unitaries of both classes are the orthogonal sum of unitary blocks,
which in turn split into non-unitary building blocks.
Long unitaries are
\begin{equation}
    \label{longdirectsum}
    P_j=\overset{M+1}{\underset{r=0}{\bigoplus}}P_j^{[r]} \ , \quad P_j^{[i,r]}=\bra{i}_jP_j^{[r]}\ket{0}_{j+M} \ , \quad 1\leq j\leq N-M \ ;
\end{equation}
whereas short unitaries are
\begin{equation}
    \label{shortdirectsum}
    P_j=\overset{N-j+1}{\underset{r=0}{\bigoplus}}\,P_j^{[r]} \ , \quad P_j^{[i,r]}=\bra{i}_{j} P_j^{[r]} \ , \quad N-M< j\leq N-1 \ .
\end{equation}
We summarize the properties of long and short unitaries in Table \ref{tableP}.
\begin{table}[H]
    \centering
    \begin{tabular}{c|c|c|c|c|c}
    $P_j^{[i,r]}$ & position & input & output & $\#$ rows $\times$ $\#$ columns & formulae  \\
        \hline 
    long   & $1\leq j\leq N-M$  & $\mathsf{H}_M^{[r]}$ & $\mathsf{H}_M^{[r-i]}$ &  $\textstyle{\binom{M}{r-i}} \times \textstyle{\binom{M}{r}} $ & (\ref{P}) \\
        \hline 
    short   & $N-M< j \leq N-1$ & $\mathsf{H}_{N-j+1}^{[r]}$ & $\mathsf{H}_{N-j}^{[r-i]}$ & $\textstyle{\binom{N-j}{r-i}} \times \textstyle{\binom{N-j+1}{r}} $ &  (\ref{Pshort}) \\
        \hline
    \end{tabular}
    \caption{\label{tableP} Features of the non-unitary building blocks $P_j^{[i,r]}$ of the unitaries of ABCs.}
\end{table}
Let us now write the exact form of the non-unitary building blocks of ABCs~\footnote{
    One may wonder if any simplification occurs in the large $N$ limit.  
    However, 
    Bethe states prepared by ABCs are not suited for addressing the thermodynamic limit.  
    Instead, the thermodynamic Bethe Ansatz \cite{Takahashi99,vanTogeren16},  
    based on the density of Bethe roots,  
    rather than the ABA or the CBA for constructing Bethe states at finite $N$,  
    is the proper approach for analyzing the thermodynamic limit of integrable models.  
    At any rate, potential connections  
    between ABCs and the thermodynamic Bethe Ansatz in the large~$N$ limit may be worth exploring. 
}.
The realization of Bethe states as the MPS of the CBA
(\ref{inhMPS}),
which provides the means to construct Bethe states
with $0\leq r\leq M$ magnons, 
enables us to compute explicit expressions.
We begin with long unitaries in Sub-subsection 
\ref{sssLong}, 
which are simpler.
In Sub-subsection 
\ref{sssShort},
we focus on short unitaries,
where we show how to eliminate the qubits post-selected 
on~$|0\rangle^{\otimes M}$ 
by defining a new short tensor in the MPS. 
Our approach to short unitaries differs from that 
of~\cite{Ruiz23}. 
We prove the equivalence between both approaches in 
Appendix~\ref{appshort}.

\subsubsection{Long Unitaries}

\label{sssLong}

Formula (\ref{Pjgauge}) implies the non-unitary building blocks of long unitaries are
\begin{equation}
    \label{P}
    P_j^{[i,r]} = X_{j+1}^{-1[r-i]} \, \Lambda_j^{[i,r]} \, X_j^{[r]} \ ,
\end{equation}
because they are unaffected by the elimination of post-selected qubits.
The matrix elements of~$\Lambda_j^{[i,r]}$ 
are 
(\ref{Lambda0r}) 
and 
(\ref{Lambda1r}).
To write the formulae of
$X_j$ 
and 
$X_j^{-1}$, 
we need Bethe states with $r$ magnons over~$k$ spins.
We choose that the states have support in the last spins of the spin chain in view of the architecture of ABCs. 
In other words,
Bethe states belong to
\begin{equation} 
    \mathsf{H}_k=\underset{\ell=j_k}{\overset{N}{\bigoplus}}\,\mathsf{h}_\ell \ ,
\end{equation}
where
\begin{equation}
   j_k:=N-k+1 \ . 
\end{equation} 
According to 
(\ref{inhMPS}),
the Bethe state with quasi-momenta 
$p_{m_a,\ell}$ 
chosen from 
$p_{1,\ell},\ldots,p_{M,\ell}$ 
is
\begin{align}
    \label{inhpartial}
    |\Psi_{k,\alpha}^{[r]} \rangle  &=
    \sum_{i_\ell=0,1}
    \bra{0}^{\otimes M}
    \Lambda_N^{i_N}\ldots \Lambda_{j_k + 1}^{i_{j_k + 1}}\Lambda_{j_k}^{i_{j_k}}
    \ket{m_1 \ldots m_r}_{M}
    \ket{i_{j_k}\ldots i_N}_k \\
    &=\underset{j_k\leq n_1<\ldots<n_r\leq N}{\sum}\,
    \underset{a_p\neq a_q}{\sum_{a_1,\ldots,a_r=1}^r}
    \left[\underset{1\leq q<p\leq r}{\prod}s_{m_{a_q} m_{a_p} }\right]
    \left[\overset{r}{\underset{p=1}{\prod}} 
    \overset{n_p-1}{\underset{\ell=j_k}{\prod}} \,\, x_{m_{a_p},\ell}\right]
    | n_1\ldots n_r \rangle_k \ . \nonumber
\end{align}
Since we focus on long unitaries, 
$1\leq j_k\leq N-M$, 
hence 
$M+1\leq k\leq N$. 
The number of magnons 
$r$ 
is always smaller than the number of spins 
$k$ 
over which they propagate.

\begin{figure}[ht]
    \centering
    \includegraphics[width=\textwidth]{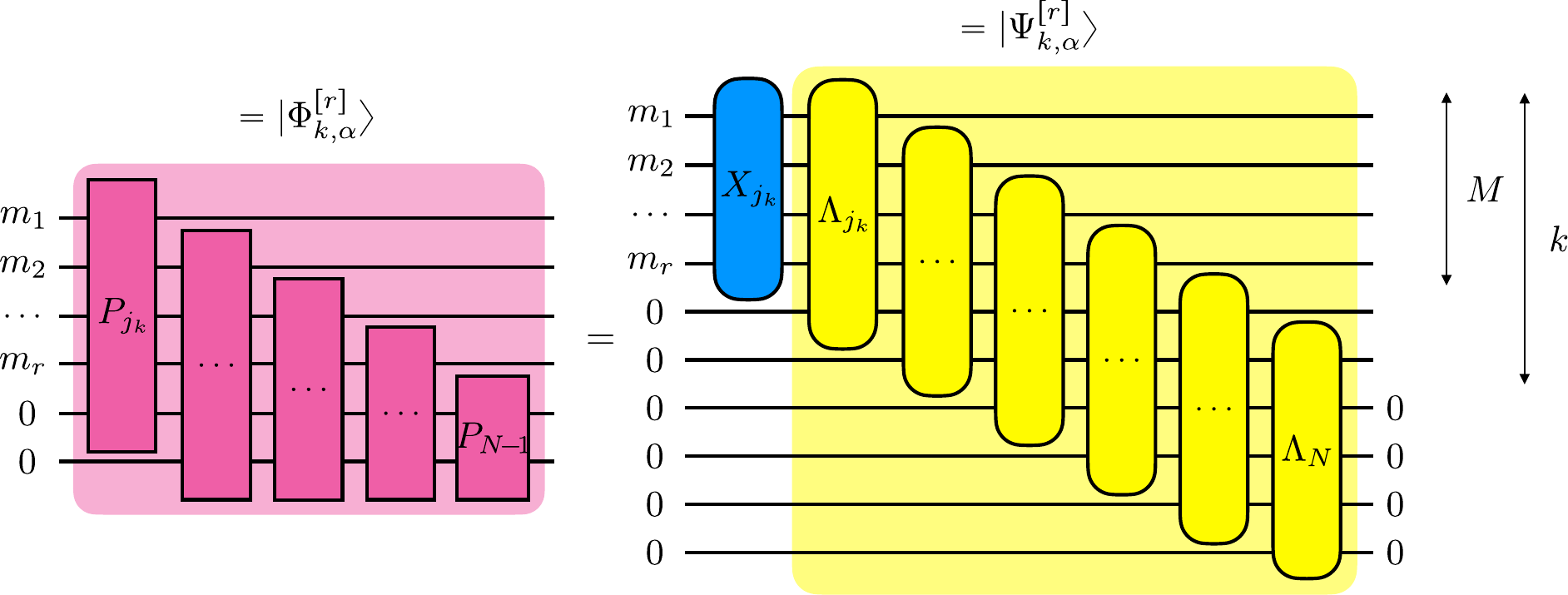}
    \caption{
    The local gauge transformation 
    $X_{j_k}$ 
    as the change-of-basis matrix that orthonormalizes the set Bethe states with 
    $r$ 
    magnons over 
    $k$
    spins.
    }
    \label{figXonLambda}
\end{figure}

Bethe states are in one-to-one correspondence with the elements of the computational basis of 
$\mathsf{H}_k^{[r]}$, as we already mentioned. 
This fact permits us to identify the MPS of the CBA with the invertible mapping
\begin{equation}
    \label{longMPS}
    \begin{matrix}
    \mathrm{MPS}_{j_k}:&\mathsf{H}_M^{[r]}&\mapsto&\mathsf{H}_k^{[r]} \\
        &|m_1 \ldots m_r\rangle_M&\mapsto&|\Psi_{k,\alpha}^{[r]}\rangle 
    \end{matrix}
    \ .
\end{equation} 
Since Bethe states are linearly independent thanks to the assumption 
$u_a\neq u_b$,
they span a linear basis of the Hilbert space of the last 
$k> M$ 
spins. 
On the other hand, 
the last $k-1$ unitaries of ABCs prepare
\begin{equation}
    |\Phi_{k,\alpha}^{[r]}\rangle =  
    P_{N-1}\ldots P_{j_k}|m_1\ldots m_r\rangle_M \ .
\end{equation}
The set 
$|\Phi_{k,\alpha}^{[r]}\rangle$ 
is an orthonormal basis of 
$\mathsf{H}_k^{[r]}$ 
because unitary transformations preserve both the orthonormality and the completeness of the basis
$\ket{m_1 \ldots m_r}_M$.
Therefore, we can think about the quantum sub-circuit of last $k-1$ unitaries as the unitary mapping
\begin{equation}
    \label{longQC}
    \begin{matrix}
    \mathrm{ABC}_{j_k}:&\mathsf{H}_M^{[r]}&\mapsto&\mathsf{H}_k^{[r]} \\
        &|m_1  \ldots m_r\rangle_M&\mapsto&|\Phi_{k,\alpha}^{[r]}\rangle 
    \end{matrix}
    \ .
\end{equation}
It follows that the matrix $X_j$ performs the change of basis
\begin{equation}
    \label{GS}
    |\Phi_{k,\alpha}^{[r]}\rangle = \sum_{\beta=1}^{\textstyle{\binom{k}{r}} }X_{j_k,\beta\alpha}^{[r]}|\Psi_{k,\beta}^{[r]}\rangle \ .
\end{equation}
We depict this formula in Figure \ref{figXonLambda}.
We note that the identification of $X_j$ with the \mbox{change-of-basis} matrix  
relies on the existence of the mapping~(\ref{longMPS}),  
which, as we showed at the end of Subsection~\ref{ssCBAhom},  
follows from the invariance of the F-basis with respect to the exchange of qubits.

We choose the change-of-basis matrix in (\ref{GS})
to correspond to the standard Gram-Schmidt process,
although it is important to notice that 
$X_j$ 
is only determined up to unitary rotations.
If we use the Gram matrix of Bethe states
\begin{equation}
\label{Cab}
    C_{k,\alpha\beta}^{[r]}
    =\langle{\Psi_{k,\alpha}^{[r]}}|{\Psi_{k,\beta}^{[r]}}\rangle\ ,
\end{equation}
the closed formulae of the matrix elements of $X_j$ are~\cite{Ruiz23}
\begin{equation}
    \label{X}
    \begin{split}
    X_{j_k,\alpha\alpha}^{[r]} &= \sqrt{\frac{\det_{\alpha-1} C_k^{[r]}}{\det_\alpha  C_k^{[r]}}} \ , \\
    X_{j_k,\alpha\beta}^{[r]} &= 0 \! \quad\!  \mathrm{if} \! \quad\! \alpha>\beta \ , \\ 
    X_{j_k,\alpha\beta}^{[r]} &= -\frac{\det_{\beta-1} C_{k,\alpha \to \beta}^{[r]}}{\sqrt{\det_{\beta-1} C_k^{[r]} \det_{\beta} C_k^{[r]}} } \quad \mathrm{if} \! \quad\! \alpha<\beta \ , 
    \end{split}
\end{equation}
where $\det_\alpha $ denotes the corner principal
$\alpha\times\alpha$-minor 
and 
$\vphantom{x}_{\alpha \to \beta}$ 
denotes the replacement of the 
$\alpha$-th 
by the
$\beta$-th 
column.
The matrix 
$X_j$ 
is upper-triangular. 
Therefore, 
$X_j^{-1}$ 
is upper-triangular as well,
and its matrix elements read~\cite{Ruiz23}
\begin{equation}
    \label{Xm1}
    X_{j_k,\alpha\beta}^{-1[r]}
     = 
    \frac{\det_\alpha C_{k, \alpha \to \beta}^{[r]}}{\sqrt{\det_{\alpha-1} C_k^{[r]} \det_{\alpha} C_k^{[r]}} } \ .
\end{equation} 
This matrix provides the Cholesky factorization of the Gram matrix (\ref{Cab}) by construction:
\begin{equation}
    \label{Cholesky}
    C_k^{[r]}=X_{j_k}^{-1[r]\dagger} {X_{j_k}^{-1[r]}} \ .
\end{equation}
Table~\ref{tableXLXm1} 
summarizes the features of the matrices in (\ref{P}).

The reason for using the Gram matrix (\ref{Cab}) 
to compute closed formulae is that we know the entries exactly,  
thanks to the knowledge of the Bethe 
states~(\ref{inhpartial}).  
However, 
this knowledge implies that neither the scalar products among Bethe states nor the associated Gram matrices can be evaluated efficiently in general.
The scalar product between two Bethe states 
with~$r$ 
magnons 
over~$k$ 
qubits involves the sum  
of~$D=\displaystyle\binom{k}{r}$ 
distinct terms.  
Given~$k$ and~$r$, 
the Gram 
matrix
contains~$(1/2)D(D+1)$ 
linearly independent scalar products.  
Gram matrices
(for long unitaries) 
must be computed for 
each~\mbox{$1\leq k\leq N-M$}
and $0\leq r\leq M$.  
The cost of these computations rapidly increases 
with~$N$ 
and~$M$ 
on a classical computer.  
Moreover, storing all the Gram matrices 
(or, 
equivalently, 
their numerical Cholesky factorization,  
which scales cubically with their size)  
demands an increasingly unfeasible amount of memory.  
These limitations must be taken into account in the classical numerical computation of the unitaries.  
Nonetheless, 
they do not diminish the usefulness of exact formulae for the unitaries in quantum computing.  
Exact formulae provide valuable data about the unitaries and,  
in particular, about their decomposition into one- and two-qubit unitaries.

\begin{table}[ht]
    \centering
    \begin{tabular}{c|c|c|c|c}
        $1\leq j\leq N-M$ & input & output & $\#$ rows $\times$ $\#$ columns & formulae \\
        \hline 
        $\Lambda_j^{[i,r]}$ & $\mathsf{H}_M^{[r]}$ & $\mathsf{H}_M^{[r-i]}$ & $\textstyle{\binom{M}{r-i}}\times \textstyle{\binom{M}{r}}$ & (\ref{Lambda0r})--(\ref{Lambda1r}) \\
        \hline 
        $X^{[r]}_j$ & $\mathsf{H}_M^{[r]}$ & $\mathsf{H}_M^{[r]}$ & $\textstyle{\binom{M}{r}} \times \textstyle{\binom{M}{r}}$ & (\ref{X}) \\
        \hline
        $X^{-1[r-i]}_{j+1}$ & $\mathsf{H}_M^{[r-i]}$ & $\mathsf{H}_M^{[r-i]}$ & $\textstyle{\binom{M}{r-i}} \times \textstyle{\binom{M}{r}}$ & (\ref{Xm1})\\
    \end{tabular}
    \caption{\label{tableXLXm1} Features of the constituents of long unitaries.}
\end{table}

It remains to demonstrate the unitarity of long unitaries 
(\ref{unitarityP}).
In the definition of~$\Lambda_j$,
we left 
$\Lambda_j|1\rangle_j$ 
unspecified because it did not appear in the MPS.
This freedom allows us to determine~$P_j|1\rangle_j$
at will, 
which we can choose at our best convenience to ensure unitarity.
On the other hand, we are left to prove that
\begin{equation}
    \bra{0}_{j_k} P_{j_k}^\dagger P_{j_k}\ket{0}_{j_k}=1_{2^{M}} \ ,
\end{equation}
which is equivalent to
\begin{equation}
    \label{unitarityLambdajk}
    C_{k}^{[r]}
    =\Lambda_{j_k}^{[0,r]\dagger}C_{k-1}^{[r]}\Lambda_{j_k}^{[0,r]}
    +\Lambda_{j_k}^{[1,r]\dagger}C_{k-1}^{[r-1]}\Lambda_{j_k}^{[1,r]} \ ,
\end{equation}
thanks to the Cholesky factorization 
(\ref{Cholesky}).
We present the proof of 
(\ref{unitarityLambdajk}) 
in Appendix 
\ref{appunitaries}.

\subsubsection{Short Unitaries}

\label{sssShort}

Our construction of short unitaries begins with the following observation.
The matrices 
$X_{j_k}$ 
of long unitaries orthonormalize the set of linearly independent Bethe states 
with \mbox{$1\leq r\leq M$}
magnons over the last
$M< k\leq N$
spins.
The computation of $X_{j_k}$ relies on that the mappings defined by the MPS 
(\ref{longMPS})
and by the quantum sub-circuit
(\ref{longQC})
are invertible and unitary, respectively.
Short unitaries cannot make use of the mapping defined by the MPS 
(\ref{longMPS}),
since it is non-invertible if $1\leq k< M$~\footnote{
    The MPS of the last~$N-M+1$ 
    tensors does not suffer from the same issue because it prepares states over~$M$ spins.  
    However, we 
    classify~$P_{N-M+1}$ 
    as a short unitary as it acts 
    on~$M$ 
    qubits after removing post-selected qubits.}.
The reason is that Bethe states with 
$1\leq r\leq k$
magnons over the last 
$1\leq k< M$ 
spins are not all linearly independent,
as they carry every possible subset of~$r$ 
quasi-momenta out of
$p_{1,j},\ldots,p_{M,j}$.
Furthermore,
Bethe states are ill-defined if the number of magnons is greater than the number of spins, 
that is, if 
$k< r\leq M$. 

\begin{figure}[ht]
    \centering
    \includegraphics[width=0.4\textwidth]{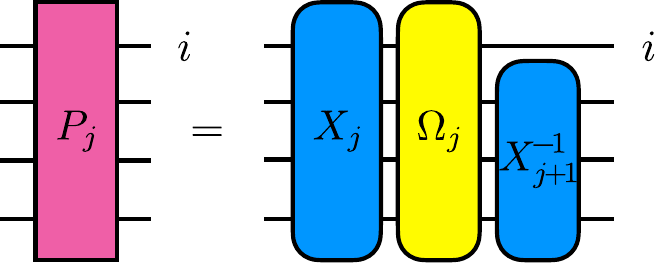}
    \caption{
    Definition of the last
    $k+1$-qubit 
    unitaries out of new non-unitary tensor of the MPS of the CBA by a local gauge transformation.
    The size of the unitaries is dictated by the number of linearly independent Bethe states over
    $k$
    spins.
    }
    \label{figPOmega}
\end{figure}

Our strategy to construct the short unitaries consists of two steps.
First, 
we replace the first~$M+1$ 
tensors 
$\Lambda_{j}$ 
of the MPS 
(\ref{inhMPS}) 
by the smaller non-unitary tensors
\begin{equation}
    \label{Omegajk}
    \Omega_{j_k}=\Omega_{j_k}(v_{j_k};u_1,\ldots,u_k)\in\mathrm{End}(\mathsf{H}_{k+1}) \ .
\end{equation}
The tensor 
$\Omega_{j_k}$
just acts on 
$k+1$ 
qubits, unlike
$\Lambda_{j_k}$. 
By definition,
the MPS of 
(\ref{Omegajk})
must only construct the maximal number of linearly independent Bethe states over 
$k$ 
spins when acting on the computational basis.
The number of magnons 
$r$ 
of Bethe states is bounded from above by 
$k$.
Second, 
we define the short unitaries by the local gauge transformation 
\begin{equation}
    \label{Pjgaugeshort}
    P_{j_k}= X_{j_k+1}^{-1} \Omega_{j_k} X_{j_k} \ , \quad N-M< j_k\leq N-1 \ ,
\end{equation}
whose matrices follow from the orthonormalization of the set of Bethe states.
We depict (\ref{Pjgaugeshort}) in Figure \ref{figPOmega}.
The new set of matrices are unitary if
\begin{equation}
    \label{PPk}
    P_{j_k}^\dagger P_{j_k} = 1_{2^{k}} \ .
\end{equation}
We stress the strategy differs from that of 
\cite{Ruiz23}, 
which is based on an Ansatz. 
Our approach is advantageous in that it enables us to demonstrate the construction of short unitaries rigorously.

\begin{figure}[ht]
    \centering
    \includegraphics[width=0.6\textwidth]{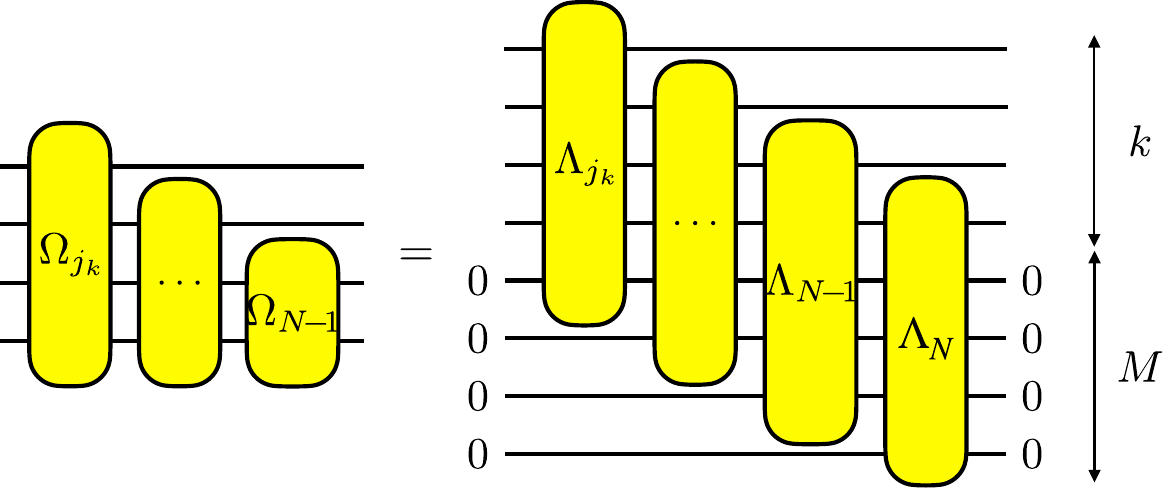}
    \caption{
    Equivalence between the MPS of the CBA based on the
    $M+1$-tensor
    (\ref{Lambdaj}) and on the~\mbox{$k+1$-tensor}
    (\ref{Omegajk}). 
    Despite the difference in the number of input qubits,
    the equivalence holds due to the linear dependence of Bethe states over 
    $k<M$ 
    spins.
    }
    \label{figOmegaLambda}
\end{figure}

The tensor (\ref{Omegajk}) breaks into blocks of definite total spin along the 
$z$-axis by assumption, which in turn split into non-unitary matrices:
\begin{equation}
    \label{Omegadecomp}
    \Omega_{j_k}=\overset{k}{\underset{r=0}{\bigoplus}}\,\,\Omega_{j_k}^{[r]} \ , 
    \quad \Omega_{j_k}^{[i,r]}=\bra{i}_{j_k}\Omega_{j_k}^{[r]} \ ,
\end{equation}
where we recall
$\bra{i}_k$ belongs to 
the Hilbert space of the $k$-th qubit.
The number of rows and columns of the non-unitary matrices is
\begin{equation}
    \#~\mathrm{rows} \times \#~\mathrm{columns~of}~\Omega_{j_k}^{[i,r]} =\binom{k-1}{r-i}\times \binom{k}{r}\ .
\end{equation}
The defining property of the new tensor is that it specifies the mapping
\begin{equation}
    \label{shortMPS}
    \begin{matrix}
    \mathrm{MPS}_{j_k}:&\mathsf{H}_k^{[r]}&\mapsto&\mathsf{H}_k^{[r]} \\
        &|m_1 \ldots m_r\rangle_k&\mapsto&|\Psi_{k,\alpha}^{[r]}\rangle = \Omega_N\ldots \Omega_{j_{k-1}}\Omega_{j_k}
        \ket{m_1 \ldots m_r}_{k}
    \end{matrix} \ ,
\end{equation} 
which is invertible. 
We stress the MPS based on 
$\Omega_j$ 
also corresponds to the CBA.
Bethe states thus computed carry 
$r$ 
quasi-momenta
$p_{m_1,\ell}\ldots,p_{m_r,\ell}$,
chosen among 
$p_{1,\ell},\ldots,p_{k,\ell}$.
We depict the equivalence between the MPS of 
$\Omega_j$ 
and 
$\Lambda_j$ 
in Figure 
\ref{figOmegaLambda}.

Let us determine the non-unitary tensor.
We begin with 
$j_k=N$, 
that is,
$k=1$.
Bethe states on the last spin are just the elements of the computational basis:
\begin{equation}
    |\Psi^{[0]}_{1,1}\rangle = |0 \rangle \ , \quad |\Psi^{[1]}_{1,1}\rangle = |1 \rangle \ .
\end{equation}
Therefore,
\begin{equation}
    \Omega_{N}^{[0]}|0\rangle=|0\rangle \ , \quad \Omega_{N}^{[1]}|1\rangle=|1\rangle \ .
\end{equation}
We deduce the last tensor is the identity matrix:
\begin{equation}
    \Omega_{N}=
    \begin{bmatrix}
        \Omega_{N}^{[0,0]} & 0 \\
        0 & \Omega_{N}^{[1,1]}
    \end{bmatrix}
    =
    \begin{bmatrix}
        1 & 0 \\
        0 & 1
    \end{bmatrix} \ .
\end{equation}
Since the last tensor is trivial, 
the 
$N$-th 
unitary is trivial as well, 
hence its absence in ABCs.

Let 
$j_k=N-1$, 
which implies 
$k=2$.
According to (\ref{inhMPS}) 
and
(\ref{shortMPS}), 
we must have
\begin{equation}
    \begin{split}
        \Omega_{N-1}^{[0]}|00\rangle&=|00\rangle \ , \\
        \quad 
        \Omega_{N-1}^{[1]}|10\rangle&=|10\rangle+x_{1,N-1}|01\rangle \ , \\
        \quad 
        \Omega_{N-1}^{[1]}|01\rangle&=|10\rangle+x_{2,N-1}|01\rangle \ , \\
        \Omega_{N-1}^{[2]}|11\rangle&=(s_{12}x_{2,N-1}-s_{21}x_{1,N-1})|11\rangle \ .
    \end{split}
\end{equation}
Therefore,
\begin{equation}
    \label{OmegaN1}
    \begin{split}
    \Omega_{N-1}\!=\!
    \begin{bmatrix}
        \Omega_{N-1}^{[0,0]} & 0 & 0 & 0 \\
        0 & \Omega_{N-1,1}^{[0,1]} & \Omega_{N-1,2}^{[0,1]} & 0 \\
        0 & \Omega_{N-1,1}^{[1,1]} & \Omega_{N-1,2}^{[1,1]} & 0 \\
        0 & 0 & 0 & \Omega_{N-1}^{[1,1]} \\
    \end{bmatrix}
    \!=\! 
    \begin{bmatrix}
        1 & 0 & 0 & 0 \\
        0 & \!\!x_{2,N-1} & \!\!x_{1,N-1} & 0 \\
        0 & 1 & 1 & 0 \\
        0 & 0 & 0 & \!\!\!\!\!\!s_{12}x_{2,N-1}\!-\!s_{21}x_{1,N-1} \\
    \end{bmatrix}
    \end{split}
     .
\end{equation}
The first non-trivial tensor allows us to compute the remainder.

The computation of the tensor for 
$N-M<j<N-1$ 
is inductive. 
We detail the derivation in Appendix
\ref{appshort}.
The result is
\begin{equation}
    \label{Omegajkentries}
    \Omega_{j_k,{\alpha\beta}}^{[i,r]} =\frac{\det B^{[r-i]}_{k-1,\alpha\rightarrow\beta}}{\det B^{[r-i]}_{k-1}} \ ,
\end{equation}
where we used the indexation of Appendix \ref{appindexation}, and
\begin{equation}
    \label{B}
    B_{k-1,\lambda\mu}^{[r]}=\vphantom{x}_{k-1}\langle \lambda |\Psi_{k-1,\mu}^{[r]} \rangle \ ; \!\quad
    \big(B_{k-1,\alpha\rightarrow\beta}^{[r-i]}\big)_{\lambda\mu}=
    \begin{cases}
    \vphantom{x}_{k-1}\langle \lambda |\Psi_{k-1,\mu}^{[r-i]} \rangle \quad &\si \quad \mu\neq \alpha \\
    \vphantom{x}_{k}\langle \lambda |\Psi_{k,\beta}^{[r]} \rangle \quad &\si \quad i=0 \ , \quad \mu=\alpha \\
    \vphantom{x}_{k}\langle 
        \lambda+\textstyle{\binom{k-1}{r}} 
        |\Psi_{k,\beta}^{[r]} \rangle \quad &\si \quad i=1 \ , \quad \mu=\alpha
    \end{cases}
    \ .
\end{equation}
The matrix 
$B^{[r]}_{k-1}$ 
performs change of basis from the computational basis to the set of Bethe states in
$\mathsf{H}_{k-1}^{[r]}$. 
The matrix 
$B_{k-1,\alpha\rightarrow\beta}^{[r]}$ 
is the result of the replacement of the 
$\alpha$-th 
column by the relevant entries of the Bethe state that the MPS
(\ref{shortMPS}) 
prepares.

The formula of the non-unitary building blocks of short unitaries is
\begin{equation}
    \label{Pshort}
    P_j^{[i,r]} = X_{j+1}^{-1[r-i]} \, \Omega_j^{[i,r]} \, X_j^{[r]} \ .
\end{equation}
The quantum sub-circuit of the last
$k-1$ 
short unitaries maps the computational basis to an orthonormal basis of 
$\mathsf{H}_k^{[r]}$:
\begin{equation}
    \begin{matrix}
    \mathrm{ABC}_{j_k}:&\mathsf{H}_k^{[r]}&\mapsto&\mathsf{H}_k^{[r]} \\
    &|m_1 \ldots m_r\rangle_k&\mapsto&|\Phi_{k,\alpha}^{[r]}\rangle = P_{N-1}\ldots P_{j_k} |m_1 \ldots m_r\rangle_k 
    \end{matrix} 
    \ .
\end{equation}
The matrix
$X_{j_k}$ 
orthonormalizes Bethe states,
as we already mentioned.
We choose the matrix to be defined by the Gram-Schmidt process again.
The entries of the matrix are
(\ref{X}) 
with a suitable adjustment in the ranges of indices.
The inverse matrix 
$X_{j_k}^{-1}$ provides the Cholesky factorization of the Gram matrix
(\ref{Cab}), 
and the entries of the inverse matrix are
(\ref{Xm1}) 
with the adjustment of the ranges of indices.
Table
\ref{tableXLXm2} 
summarizes the features of the matrices in 
(\ref{Pshort}).

\begin{table}[ht]
    \centering
    \begin{tabular}{c|c|c|c|c}
        $N-M< j_k\leq N-1$ & input & output & $\#$ rows $\times$ $\#$ columns & formulae \\
        \hline 
        $\Omega_{j_k}^{[i,r]}$ & $\mathsf{H}_{k}^{[r]}$ & $\mathsf{H}_{k-1}^{[r-i]}$ & $\textstyle{\binom{k-1}{r-i}} \times \textstyle{\binom{k}{r}} $ & (\ref{Omegajkentries})--(\ref{B}) \\
        \hline 
        $X^{[r]}_{j_k}$ & $\mathsf{H}_{k}^{[r]}$ & $\mathsf{H}_{k}^{[r]}$ & $\textstyle{\binom{k}{r}} \times \textstyle{\binom{k}{r}} $ & (\ref{X}) \\
        \hline 
        $X^{[r-i]-1}_{j_k+1}$ & $\mathsf{H}_{k-1}^{[r-i]}$ & $\mathsf{H}_{k-1}^{[r-i]}$ & $\textstyle{\binom{k-1}{r-i}} \times \textstyle{\binom{k-1}{r-i}} $ & (\ref{Xm1}) \\
    \end{tabular}
    \caption{\label{tableXLXm2} Features of the constituents of short unitaries.}
\end{table}

The last step of the construction of short unitaries is the proof of 
(\ref{PPk}). 
According 
to~(\ref{Pshort})
and the Cholesky factorization 
(\ref{Cholesky}),
unitarity is equivalent to
\begin{equation}
    \label{unitarityOmegajk}
    C_{k}^{[r]}
    =\Omega_{j_k}^{[0,r]\dagger}C_{k-1}^{[r]}\Omega_{j_k}^{[0,r]}
    +\Omega_{j_k}^{[1,r]\dagger}C_{k-1}^{[r-1]}\Omega_{j_k}^{[1,r]} \ .
\end{equation}
We present the demonstration of 
(\ref{unitarityOmegajk}) 
in Appendix 
\ref{appunitaries}.

\section{Conclusions}

\label{sconclusions}

In this work, 
we systematized the ABCs of~\cite{Sopena22,Ruiz23},
a recent proposal of quantum circuits that prepare arbitrary Bethe states of the standard spin-$1/2$ XXZ model with periodic boundaries. 
We demonstrated that the exact unitaries from \cite{Ruiz23} can alternatively be obtained by performing a change of basis in the auxiliary space of the ABA. 
The resulting basis is equivalent to the F-basis, known from the theory of quantum-integrable models.

The key property of the F-basis is that it is symmetric with respect to the exchange of qubits. 
When applied in the auxiliary space of the ABA, 
the resulting MPS is invariant with respect to the exchange of the ancillae. 
The explicit wave functions of this MPS are scattering plane waves, 
thus establishing a natural connection to the CBA. 
As a by-product, the same MPS generates Bethe states with an arbitrary number of magnons simply by changing the initialization in the auxiliary space. 
While it is relatively straightforward to prove this, we observed that it had not yet been discussed in the F-basis literature.
Furthermore, 
the symmetry of the F-basis clarified how to rigorously eliminate the auxiliary space in the final circuits, 
so that the ABCs have no post-selected qubits.
We showcased our approach with new circuits for preparing the exact Bethe states of the 
inhomogeneous spin-$1/2$ XXZ model with periodic boundaries.
We believe that the symmetry with respect to the exchange of qubits in the auxiliary space holds potential for constructing circuits for other quantum-integrable models 
and could, 
hopefully, 
aid in identifying efficient decompositions.
Our results open up promising avenues for future research, some of which we briefly discuss below.

First, our ABCs could be applied to models closely related to the inhomogeneous 
spin-$1/2$ 
XXZ model. 
A straightforward idea is to consider the staggered spin chain \cite{Frahm13}. 
This model can be obtained by choosing alternating inhomogeneities 
$v_{2j-1}=-\im v$ 
and $v_{2j}=\im v$.
The Hamiltonian is composed of two- and three-body densities and, 
much like the Hamiltonian of the homogeneous spin chain, exhibits a rich phase diagram. 
This model also holds relevance from another perspective: 
the alternating spin chain is employed in the ``integrable Trotterization'' of the XXZ model \cite{Vanicat17}. 
This implies that our circuits can construct the exact eigenstates of selected integrable quantum circuits, 
which are utilized for simulating non-equilibrium dynamics in discrete time. 
Another generalization of our circuits  
involves the preparation of Bethe states in spin-$s$ XXZ models~\cite{Korepin93,Faddeev96},  
which possess an F-basis in both the quantum and auxiliary spaces~\cite{Terras99}.  
The main distinction in these models is that the spins are spin-$s$ qudits,  
while the ancillae remain qubits.  
Although the corresponding MPS tensor of the CBA is available and retains symmetry under the exchange of ancillae,  
care must be taken in constructing a quantum circuit, particularly in determining  
the change-of-basis matrices of the gauge transformation,  
since the tensors act on spin-$1/2$ qubits but realize Bethe wave functions for spin-$s$ qudits.  
This mismatch, in particular, could obstruct the elimination of post-selected qubits in the final circuit.

Another spin system that is worth considering is the
Richardson-Gaudin (RG) model \cite{Gaudin14,Richardson77}, which realizes doubly degenerate fermions with 
pair-wise interaction and is integrable~\cite{Cambiaggio97}.
The eigenstates of the RG model are Bardeen-Cooper-Schrieffer states,
which can be computed using the ABA \cite{Amico00,Amico01}.
The isotropic and anisotropic RG models are related to the ``quasi-classical'' limits of the transfer matrices of the inhomogeneous 
XXX~\cite{Amico00,vonDelft02} 
and XXZ 
models~\cite{Amico01,vonDelft02}, 
respectively;
therefore,
we expect a similar limit to be applicable to the unitaries of ABCs.
The RG model requires to deform the periodic boundary conditions by a diagonal twist,
which adds a new layer of complexity to the method without precluding its applicability~\cite{workinprogress}.

A more challenging task is the construction of quantum circuits for other spin chains which can be solved by the Bethe Ansatz. 
One potential candidate is the solid-on-solid model equivalent to the inhomogeneous spin-$1/2$ XYZ model \cite{Baxter72},
which provides the means to construct Bethe states of this completely anisotropic spin chain \cite{Takhtadzhan79}. 
This ice-type model has an 
F-basis~\cite{Albert00ii}, 
but it is linked to the dynamical YBE~\cite{Felder94} rather than the standard YBE~(\ref{YBdiff}),
thus posing a new challenge to ABCs.
Even more complex are higher-rank spin chains. 
While these models do have an F-basis \cite{Albert00,Yang05,McAteer12}, 
it remains uncertain whether it would be beneficial for constructing nested Bethe states. 
An F-basis in the auxiliary space would be necessary, but achieving such a generalization remains unclear.
Ultimately, the aim is to construct the nested Bethe states, with the nesting occurring in the auxiliary space. 
Perhaps the more recent methods discussed in~\cite{Gromov16,Liashyk18} for the ABA in these models could prove advantageous.

\section*{Acknowledgements}

R.R. is grateful to Juan Miguel Nieto García for useful discussions.
The work of E.L., 
R.R., 
G.S.,
and
A.S. 
has been financially supported by the Spanish Agencia Estatal de Investigaci{\'o}n through 
``Instituto de Física Teórica Centro de Excelencia Severo Ochoa CEX2020-001007-S'' 
and 
PID2021-127726NB-I00 funded by MCIN/AEI/10.13039/501100011033, 
by European Regional Development Fund, 
and the ``Centro Superior de Investigaciones Científicas Research Platform on Quantum Technologies PTI-001'', 
by the Ministerio de Economía, Comercio y Empresa through the Estrategia Nacional de Inteligencia Artificial project call ``Quantum Spain'', 
and by the European Union through the ``Recovery, Transformation and Resilience Plan - NextGenerationEU'' within the framework of the ``Digital Spain 2025 Agenda''.
R. R. is supported by the Universidad Complutense de Madrid, 
Ministerio de Universidades, and the European Union - NextGenerationEU through contract CT18/22.
B. P. is supported by the NKFIH excellence grant TKP2021-NKTA-64.
The authors are grateful to the organizers of 
\textit{Exactly Solved Models and Quantum Computing} 
at Lorenz Center for support and stimulating environment while this work was being completed.
R.R. is grateful to the organizers of 
\textit{Integrable Techniques in Theoretical Physics} 
at Physikzentrum Bad Honnef for support and stimulating environment 
and
\textit{Recent Advances in Quantum Integrable Systems} 
at Laboratoire d'Annecy-le-Vieux de Physique Théorique
for stimulating environment while this work was being completed.

\appendix

\section{The Matrix-Product State of the Coordinate Bethe Ansatz}

\label{appMPSCBA}

In this appendix, 
we provide the proofs of claims about the MPS of the CBA we made in the main text.
In Subsection
\ref{appMPSCBAfirst},
we show the equivalence between the MPS of the CBA of the homogeneous spin chain 
(\ref{MPS}) 
and the MPS of 
\cite{Ruiz23}.
In Subsection
\ref{appdem},
we demonstrate the equality between both realizations of Bethe states of the inhomogeneous spin chain
(\ref{inhMPS}):
the MPS and the superposition of plane waves.
In Subsection
\ref{appinhCBA},
we prove the equivalence 
between~(\ref{inhMPS}) 
and the Bethe wave functions of
\cite{Ovchinnikov10}.

\subsection{Equivalent Matrix-Product States}

The Bethe state of 
\cite{Ruiz23} 
equals an MPS of the CBA whose tensor appears in
(54)--(57) 
therein.
The MPS of 
\cite{Ruiz23} 
follows from
(\ref{MPS})
under the replacement of the scattering amplitudes like
\begin{equation}
    \label{sab}
    s_{ab}\mapsto\frac{1}{1+x_ax_b-2\Delta x_a}=
    \frac{\sinh(u_a+\im\gamma)\sinh(u_b+\im\gamma)}{\sinh(\im\gamma)\sinh(u_a-u_b+\im\gamma)} \ ,
\end{equation}
and the following change of normalization:
\begin{equation}
    |\Psi^{[M]}_N\rangle\mapsto\left[\underset{1\leq b< a\leq M}{\prod}\frac{1}{s_{ba}s_{ab}}\right] |\Psi^{[M]}_N\rangle \ .
\end{equation}
The Bethe states of 
\cite{Ruiz23} 
are also explicitly proportional to 
(\ref{MPS}).
The change of the normalization that accounts for the equivalence is 
\begin{equation}
    |\Psi_{N}^{[M]} \rangle  \mapsto
    \left[\prod_{1\leq b<a\leq M}\frac{
        \sinh(\im\gamma)
        \sinh(u_a-u_b+\im\gamma)
        \sinh(u_b-u_a+\im\gamma)}{
        \sinh(u_a+\im\gamma)
        \sinh(u_b+\im\gamma)
        \sinh(u_a-u_b)}\right] |\Psi_{N}^{[M]} \rangle \ .
\end{equation}

\label{appMPSCBAfirst}

\subsection{Inhomogeneous Coordinate Bethe Ansatz}

\label{appdem}

Let us prove the equality between the first and second lines of
(\ref{inhMPS}).
We follow the analogous proof of Appendix B of 
\cite{Ruiz23}.
We begin with 
$N=1$. 
Formula
(\ref{Lambdaj01}) 
lead us to
\begin{equation}
    \label{MPSN1}
    \sum_{i=0,1}
        \Lambda_1^i\,
    |1\rangle^{\otimes M}
    |i\rangle
    =
    \Bigg[
        \prod_{a=1}^{M}x_{a,1}
    \Bigg]
        |1\rangle^{\otimes M}
    |0\rangle
    +
    \sum_{a=1}^{M}
    \Bigg[
        \underset{b\neq a}{\prod_{b=1}^M}\,
        s_{ab}\,
        x_{b,1}
    \Bigg]
        \sigma_a^{-}
        |1\rangle^{\otimes M}
    |1\rangle \ ,
\end{equation}
where we borrowed the notation from
(\ref{scattering}) 
and 
(\ref{xaj}). 
The rightmost state in
(\ref{MPSN1}) 
corresponds to the first spin and belongs to the quantum space.
The tensor
$\Lambda_1$ 
either preserves the ancillae on 
$|1\rangle$ 
and the spin on
$|0\rangle$
or flips the state of one single ancilla at a time in exchange for flipping the state of the spin.
The fact follows from the commutativity of
$\Lambda_1$ 
with the total spin along the 
$z$-axis over the 
$M$ 
ancillae and the spin.

If 
$N=2$, 
we obtain
\begin{equation}
    \label{MPSN2}
    \begin{split}
    \sum_{i_j=0,1}
    \Lambda_2^{i_2}\Lambda_1^{i_1}\,
    |1\rangle^{\otimes M}
    |i_1 i_2\rangle
    =
    \Bigg[\vphantom{\prod_x}
        \prod_{a=1}^{M}
        x_{a,1}\,
        x_{a,2}
    \Bigg]
        |1\rangle^{\otimes M}|00\rangle
    +
    \sum_{a=1}^{M}
    \Bigg[
        \underset{b\neq a}{\prod_{b=1}^M}\,
        s_{ab}\,
        x_{b,1}\,
        x_{b,2}
    \Bigg]
        \sigma_a^{-}
        |1\rangle^{\otimes M}
    |10\rangle\\
    +\sum_{a=1}^{M}
    \!
    \Bigg[x_{a,1}
        \underset{b\neq a}{\prod_{b=1}^M}\,
        s_{ab}\,
        x_{b,1}\,
        x_{b,2}
    \Bigg]
    \sigma_a^{-}
        |1\rangle^{\otimes M}
    |01\rangle
    +\!\!\sum_{a=1}^{M}
    \underset{c\neq a}{\sum_{c=1}^{M}}
    \Bigg[
        \underset{b\neq a}{\prod_{b=1}^M}
        \,s_{ab}
        \,x_{b,1}\!
        \underset{d\neq a,c}{\prod_{c=1}^M}
        \!
        s_{cd}
        \,x_{d,2}
    \Bigg]
    \sigma_c^{-}\!\sigma_a^{-}|1\rangle^{\otimes M}|11\rangle\ .
    \end{split}
\end{equation}
The pattern is now clear. 
If the tensor keeps the
$j$-th 
spin on 
$|0\rangle$,
it yields the product of the quasi-momentum variables 
$x_{a,j}$ 
of the ancillae that remain on 
$|1\rangle$.
If the tensor flips the state of the
$j$-th 
spin into 
$|1\rangle$,
the state of one ancilla becomes
$\ket{0}$.
Let the ancilla be at the
$a$-th 
position.
The tensor yields the product of scattering amplitudes
$s_{ab}$
and quasi-momentum variables
$x_{b,j}$
of the ancillae on 
$|1\rangle$.
The eventual projection onto 
$\bra{0}^{\otimes M}$
in the auxiliary space forces 
$M$ 
out of~$N$
spins to be on 
$|1\rangle$.
The projection and the pattern just explained lead to 
(\ref{inhMPS}).

\subsection{Equivalence with
Ovchinnikov Bethe states}

\label{appinhCBA}

We now demonstrate the equivalence between 
(\ref{inhMPS}) 
and the Bethe states of 
\cite{Ovchinnikov10}.
We must perform the non-local gauge transformation of the tensor 
(\ref{Lambdaj}):
\begin{equation}
    \Lambda_j^{i}\mapsto W_{j}^{-1}\Lambda_j^{i}W_{j-1} \ ,
\end{equation}
where
\begin{equation}
    \label{Wj}
    W_j
    ={\overset{M}{\underset{a=1}{\bigotimes}}}
    \begin{bmatrix}
        \left[\prod_{k=1}^{j-1}x_{a,k}\right]
        \prod_{b=1,b\neq a}^M s_{ab}s_{ba} & 0 \\
        0 &  
        \left[\prod_{k=j+2}^{N}x_{a,k}\right]g_{aj+1}
    \end{bmatrix}
    \ ,
\end{equation}
and we borrowed the notation from 
(\ref{scattering}) 
and 
(\ref{xaj}).
The non-locality of the transformation refers to the dependence of the 
$j$-th 
matrix 
$W_j$ 
on the quasi-momenta $p_{a,j}$ on every position of the spin.
(The product by definition equals one if the upper endpoint is smaller than the lower or vice versa.)
If we normalize
\begin{equation}
    |\Psi_{N}^{[M]}\rangle\mapsto \left[\prod_{a=1}^M\frac{f_{a1}}{g_{a1}f_{aN}}\underset{b\neq a}{\prod_{b=1}^M} \, s_{ab}s_{ba}\right] |\Psi_{N}^{[M]}\rangle \ ,
\end{equation}
to cancel the multiplicative factor induced by the transformation and follow the steps of Subsection \ref{appdem}, 
we obtain
\begin{equation}
    \label{Ovchinnikov}
    |\Psi_{N}^{[M]}\rangle =\underset{1\leq n_1<\ldots<n_M\leq N}{\sum}\
    \underset{a_p\neq a_q}{\sum_{a_1,\ldots,a_M=1}^M}
    \left[\underset{1\leq q<p\leq M}{\prod}\frac{1}{s_{a_p a_q}}\right]
    \left[\vphantom{\underset{1\leq a<b\leq M}{\prod}}
    \overset{M}{\underset{p=1}{\prod}}\,\,
    g_{a_p,n_p}\!
    \overset{N}{\underset{j=n_p+1}{\prod}}\! x_{a_p,j}\right]\!
    | n_1\ldots n_M \rangle_N \ .
\end{equation}
Formula 
(\ref{Ovchinnikov}) 
matches expressions (13) and (14) of Bethe states in \cite{Ovchinnikov10}.

\section{Indices of Algebraic Bethe Circuits}

\label{appindexation}

In this appendix, 
we define the collective indices for the ordered strings of integers that labels the computational basis of
$\mathsf{H}_k^{[r]}$.
The labeling of Subsection 3.1 of 
\cite{Ruiz23} 
is the opposite to ours.

Let 
$1\leq m_1<\ldots<m_r\leq k$
be the string that labels the state of the computational 
basis~$\ket{i_1 \ldots i_k}=\ket{m_1 \ldots m_r}_k$
of 
$\mathsf{H}_k^{[r]}$.
We define the collective index following three steps.
First,
we rephrase the string in the binary basis as a number in the decimal basis:
\begin{equation}
    \label{decimal}
    \chi:=\sum_{j=1}^{k}2^{k-j}i_j=\sum_{j=1}^{k}2^{k-j}\sum_{p=1}^r\delta_{m_p}^j\ .
\end{equation}
Note the most significant bit in the string is the first,  
then the second, 
and so on.
Next,
we arrange the numbers in the totally ordered set 
\begin{equation}
    \label{v}
    S=\left(\left\{\chi=\sum_{j=1}^{k}2^{k-j}\sum_{p=1}^r\delta_{m_p}^j : 1\leq m_1<\ldots<m_r\leq k \right\},<\right) \ , 
\end{equation}
where the order relation
$<$ 
is the standard inequality among integer. 
Finally, we assign 
$\alpha$ to the $\alpha$-th element of $S$:
\begin{equation}
    \label{alpha}
    |\alpha\rangle_{k}=| S_\alpha\rangle_{k} \ , \quad \alpha=1,\ldots,\binom{k}{r} \ .
\end{equation}
We illustrate the assignation of collective indices in Table \ref{tableindices}.
\begin{table}[H]
    \centering
    \begin{tabular}{cccc|cc|c|c}
        $i_1$ & $i_2$ & $i_3$ & $i_4$ & $m_1$ & $m_2$ & $\;\;\chi \;\; $& \; $\alpha$ \; \\
        \hline 
        0 & 0 & 1 & 1 & 3 & 4 & 3 & 1  \\
        0 & 1 & 0 & 1 & 2 & 4 & 5 & 2  \\
        0 & 1 & 1 & 0 & 2 & 3 & 6 & 3  \\
        1 & 0 & 0 & 1 & 1 & 4 & 9 & 4  \\
        1 & 0 & 1 & 0 & 1 & 3 & 10 & 5  \\
        1 & 1 & 0 & 0 & 1 & 2 & 12 & 6  \\
    \end{tabular}
    \caption{\label{tableindices} Example of assignation of collective indices for the computational basis.}
\end{table}

\section{The Short Tensor}

\label{appshort}

This appendix is devoted to demonstrations around the non-unitary tensor of short unitaries 
$\Omega_j$ 
in (\ref{Omegajk}).
In Subsection
\ref{appshortderivation}, 
we derive the formula 
(\ref{Omegajkentries}) 
for the entries.
In Subsection 
\ref{appequivABC},
we prove the equivalence with the Ansatz of
\cite{Ruiz23}.

\subsection{Closed Formulae}

\label{appshortderivation}

We derive
(\ref{Omegajkentries}) 
by induction.
The base of the induction is (\ref{OmegaN1}),
for which
$j=N-1$.
Formula~(\ref{Omegajkentries}) 
directly holds for 
(\ref{OmegaN1}).
We must prove the inductive step 
for~\mbox{$N-M<j<N-1$}. 
It follows from 
(\ref{shortMPS}) 
that
\begin{align}
    &|\Psi^{[r]}_{k,\alpha}\rangle
    =
    \Omega_{N-1}\ldots\Omega_{j_{k-1}}\Omega_{j_k}|m_1 \ldots m_r \rangle_{k}
    \\
    &=
    \sum_{i=0,1}
    \,
    \sum_{1\leq n_1<\ldots <n_{r-i}\leq k-1}
    |i\rangle\!
    \left[
    \vphantom{\sum_{1\leq n_1<\ldots <n_{r-i}<k-1}}    
    \Omega_{N-1}
    \ldots 
    \Omega_{j_{k-1}}
    |{n_1\ldots n_{r-i}}\rangle_{\,k-1}
    \right]\!\!
    \vphantom{x}_{\,k-1}\bra{n_1\ldots n_{r-i}}
    \Omega_{j_k}^{[i,r]}
    |{m_1\ldots m_r}\rangle_{\,k}
    \nonumber 
    \\
    &=|0\rangle\!\!\!
    \sum_{\beta=1}^{\textstyle{\binom{k-1}{r}} }
    \!\!\!
    |\Psi^{[r]}_{k-1,\beta}\rangle
    \,\Omega_{j_k,\beta\alpha}^{[0,r]}
    +
    |1\rangle\!\!\!
    \sum_{\beta=1}^{\textstyle{\binom{k-1}{r-1}} }
    \!\!\!
    |\Psi^{[r-1]}_{k-1,\beta}\rangle
    \,\Omega_{j_k,\beta\alpha}^{[1,r]}\ . \nonumber
\end{align}
Since Bethe states with different number of magnons are linearly independent,
two decoupled linear systems for the entries of the tensor (\ref{Omegajk}) arise, 
namely,
\begin{align}
    \label{linear}
    \begin{split}
    \sum_{\beta=1}^{\textstyle{\binom{k-1}{r}}}\!\!\!
    \vphantom{x}_{k-1}\langle \lambda|\Psi^{[r]}_{k-1,\beta}\rangle
    \,\Omega_{j_k,\beta\alpha}^{[0,r]}
    &= \vphantom{x}_k\langle \lambda|\Psi^{[r]}_{k,\alpha}\rangle \ , \quad
    \lambda = 1,\ldots,\binom{k-1}{r} \ ,
    \\
    \sum_{\beta=1}^{\textstyle{\binom{k-1}{r-1}}}\!\!\!
    \vphantom{x}_{k-1}\langle \lambda|\Psi^{[r-1]}_{k-1,\beta}\rangle
    \,\Omega_{j_k,\beta\alpha}^{[1,r]}
    &=\vphantom{x}_k\langle \lambda+{\textstyle{\binom{k-1}{r}}}\vphantom{\frac{x}{x}}|\Psi^{[r]}_{k,\alpha}\rangle \ , \quad 
    \lambda = 1,\ldots,\binom{k-1}{r-1}\ ,
    \end{split}
\end{align}
where we used the indexation of the computational basis of Appendix 
\ref{appindexation}.
The Cramer rule provides the solution for both linear systems 
(\ref{linear}) 
in terms of the change-of-basis matrices between the computational basis and Bethe states 
(\ref{B}). 
The result is
(\ref{Omegajkentries}).

\subsection{Equality with Other Short Tensors}

\label{appequivABC}

Reference 
\cite{Ruiz23} 
used another tensor to build short unitaries.
Let us demonstrate both tensors are equal.
The spin chain of
\cite{Ruiz23} 
is homogeneous,
hence we set $v_j=0$.
According to 
(54)--(57) 
and~(86) 
of 
\cite{Ruiz23}, 
we must prove
\begin{equation}
    \label{OmegaLambda}
    \Omega_{j_k,\alpha\beta}^{[i,r]}=\sum_{\lambda=1}^{\textstyle{\binom{k}{r-i}}}L_{j_{k-1},\alpha\lambda}^{[r-i]}\,\Lambda_{j_k,\lambda\beta}^{[i,r]} \ , \quad \alpha=1,\ldots,\binom{k-1}{r-i} \ , \quad \beta=1,\ldots,\binom{k}{r} \ ,
\end{equation}
where we used the indexation of Appendix~\ref{appindexation}.
The entries in left-hand side are (\ref{Omegajkentries}).
The definition of the tensor 
$\Lambda_{j_k}^i$ in the right-hand side
is
(\ref{Lambda0r})
and 
(\ref{Lambda1r})
under the adaptation of the ranges of the indices.
The MPS of
$\Lambda_{j_k}^i$ 
thus defined prepares Bethe states over 
$k$ 
spins and 
$r$
magnons with quasi-momenta in the set 
$p_{1},\ldots,p_k$.
On the other hand,
\begin{equation}
    \label{L}
    L_{j_{k-1},\alpha\beta}^{[r-i]}=\frac{\det C_{k-1,\alpha\rightarrow\beta}^{[r-i]}}{\det C_{k-1}^{[r-i]}} \ .
\end{equation}
which is rectangular:
\begin{equation}
    \#~\mathrm{rows} \times \#~\mathrm{columns}~\mathrm{of}~L_{j_{k-1}}^{[r-i]} = \binom{k-1}{r-i} \times \binom{k}{r-i} \ .
\end{equation}
The matrix $C_{k}^{[r]}$ is the Gram matrix of Bethe states (\ref{Cab}). 
However, the range of $\beta$ in $C_{k,\alpha\rightarrow\beta}^{[r]}$ 
extends beyond the scalar products of the maximal set of linearly independent Bethe states 
in~$\mathsf{H}_{k}^{[r]}$, 
with $r$ quasi-momenta among $p_1, \ldots, p_k$. 
It also includes the scalar products between this set and other linearly dependent Bethe states, 
in particular those carrying one additional quasi-momentum $p_{k+1}$ and $r-1$ quasi-momenta from $p_1, \ldots, p_k$.
The matrix (\ref{L}) is rectangular for this reason.

The first step of the demonstration of (\ref{OmegaLambda}) is 
\begin{equation}
    C_{k,\alpha\beta}^{[r]}=
    \langle\Psi_{k,\alpha}^{[r]}|\Psi_{k,\beta}^{[r]}\rangle
    =\sum_{\lambda=1}^{\textstyle{\binom{k}{r}}}
    \langle\Psi_{k,\alpha}^{[r]}\ket{\lambda}\!\vphantom{x}_k\vphantom{x}_k\!\bra{\lambda}\Psi_{k,\beta}^{[r]}\rangle
    \ .
\end{equation}
The change-of-basis matrix between the computational basis and Bethe states (\ref{B}) Cholesky-factorizes the Gram matrix (\ref{Cab}):
\begin{equation}
    \label{CBB}
    C_{k}^{[r]}=B_{k}^{[r]\dagger}B_{k}^{[r]} \ .
\end{equation}
Therefore,
\begin{equation}
    \label{LB}
    L_{j_{k-1},\alpha\beta}^{[r-i]}=
    \frac{\det \widetilde{B}^{[r-i]}_{k-1,\alpha\rightarrow\beta}}{\det B_{k-1}^{[r-i]}} \ ,
\end{equation}
where
\begin{equation}
    \big(\widetilde{B}_{k,\alpha\rightarrow\beta}^{[r]}\big)_{\lambda\mu}
    =
    \begin{cases}
    \vphantom{x}_{k}\langle \lambda |\Psi_{k,\mu}^{[r]}\rangle= B_{k,\lambda\mu}^{[r]} \quad &\si \quad \mu\neq \alpha \\
    \vphantom{x}_{k}\langle \lambda |\Psi_{k,\beta}^{[r]} \rangle \quad &\si \quad \mu=\alpha \\
    \end{cases}
    \ .
\end{equation}
The second part of the proof follows from the recurrence relation among Bethe states.

Let $i=0$. 
We use the notation
(\ref{alphaequivm}) 
in addition to
\begin{equation}
    |\lambda\rangle_{k}= |\ell_1 \ldots \ell_r\rangle_{k} \ , \quad j_k\leq \ell_1<\ldots<\ell_r\leq N \ ,
\end{equation}
following Appendix 
\ref{appindexation}.
The labeling also holds with 
$k$ 
replaced by 
$k-1$.
We apply
(\ref{Lambda0r})
and~(\ref{Omegajkentries})
to 
(\ref{OmegaLambda})
and obtain
\begin{equation}
    \label{BB}
    \det B^{[r]}_{k-1,\alpha\rightarrow\beta}
    =\det\widetilde{B}^{[r]}_{k-1,\alpha\rightarrow\beta}\left[\prod_{p=1}^{r}x_{n_p}\right] \ .
\end{equation}
Formula~(\ref{inhpartial})
with 
$v_j=0$
implies
\begin{equation}
    \vphantom{x}_k\langle\lambda|\Psi^{[r]}_{k,\beta}\rangle
    =
    \underset{a_p\neq a_q}{\sum_{a_1,\ldots,a_M=1}^r}
    \left[\underset{1\leq q<p\leq M}{\prod}s_{n_{a_q} n_{a_p} }\right]
    \left[\overset{M}{\underset{p=1}{\prod}}\,\,x_{n_{a_p}}^{\ell_p-N+k-1}\right]
    =
    \left[\prod_{p=1}^{r}x_{n_p}\right]
    \vphantom{x}_{k-1}\langle\lambda|\Psi^{[r]}_{k-1,\beta}\rangle \ ,
\end{equation}
where we took into account that 
\begin{equation}
    N-k+1<\ell_1<\ldots<\ell_r\leq N \quad \si \quad \lambda=1,\ldots,\binom{k-1}{r} \ .
\end{equation}
Formula (\ref{BB}) then follows from then multi-linearity of determinants.

Let $i=1$. 
We use the notation
(\ref{alphar})
and
\begin{equation}
    |\lambda_1\rangle_{k-1}= |\ell_2 \ldots \ell_{r}\rangle_{k-1} \ , \quad j_{k-1}\leq \ell_2<\ldots<\ell_r\leq N \ .
\end{equation}
We apply
(\ref{Lambda1r})
and
(\ref{Omegajkentries})
to
(\ref{OmegaLambda})
and obtain
\begin{equation}
    \label{BB2}
    \det B^{[r-1]}_{k-1,\alpha\rightarrow\beta}
    =
    \sum_{p=1}^{r}
    \Bigg[
        \underset{q\neq p}{\overset{r}{\underset{q=1}{\prod}}}\, 
        s_{n_p n_q}x_{n_q} 
    \Bigg]
    \det\widetilde{B}^{[r-1]}_{k-1,\alpha\rightarrow\beta_p} \ .
\end{equation}
We deduce from
(\ref{inhpartial})
with 
$v_j=0$
that
\begin{equation}
    \vphantom{x}_k\langle\lambda|\Psi^{[r]}_{k,\beta}\rangle
    =
    \sum_{p=1}^{r}
    \Bigg[
        \underset{q\neq p}{\overset{r}{\underset{q=1}{\prod}}}\, 
        s_{n_p n_q}x_{n_q} 
    \Bigg]
    \vphantom{x}_{k-1}\langle\lambda_1|\Psi^{[r-1]}_{k-1,\beta_p}\rangle \ ,
\end{equation}
where we used
\begin{equation}
        N-k+1=\ell_1<\ell_2<\ldots<\ell_r\leq N 
        \quad \si \quad 
        \lambda=\binom{k-1}{r}+1,\ldots,\binom{k}{r} \ .
\end{equation}
Multi-linearity of determinants implies (\ref{BB2}).
The proof of (\ref{OmegaLambda}) is complete.

\section{Unitarity of Algebraic Bethe Circuits}

\label{appunitaries}

This appendix is devoted to the demonstration of unitarity of ABCs. 
In Subsection 
\ref{applong},
we demonstrate 
(\ref{unitarityLambdajk}),
which implies the unitarity of long unitaries.
In Subsection
\ref{appshortunitary}, 
we demonstrate 
(\ref{unitarityOmegajk}), 
which implies the unitarity of short unitaries.

\subsection{Proof of Unitarity of Long Unitaries}

\label{applong}

The proof of 
(\ref{unitarityLambdajk}) 
amounts to a direct computation.
The entries of 
(\ref{unitarityLambdajk})
are
\begin{equation}
    \label{ulong}
    C_{k,\alpha\beta}^{[r]}=
    \Bigg[
        \overset{r}{\underset{p=1}{\prod}}\,\,
        \bar{x}_{m_p,j_k}x_{n_p,j_k} 
    \Bigg]C_{k-1,\alpha\beta}^{[r]} +
    \sum_{p,q=1}^r
    \Bigg[
        \underset{p^\prime\neq p}{\overset{r}{\underset{p^\prime=1}{\prod}}}\, 
        \bar{s}_{m_{p^\prime} m_p}\bar{x}_{m_{p^\prime},j_k} 
    \Bigg] 
    \Bigg[
        \underset{q^\prime \neq q}{\overset{r}{\underset{q=1}{\prod}}}\, 
        s_{n_{q^\prime} n_q}x_{n_{q^\prime},j_k} 
    \Bigg]C_{k-1,\alpha_{p}\beta_q}^{[r-1]} \ ,
\end{equation}
where we labeled Gram matrices by 
(\ref{alphaequivm}) 
and 
(\ref{alphar}).
Formula (\ref{ulong}) is the consequence of
\begin{align}
        C_{k,\alpha\beta}^{[r]}&=
        \sum_{j_k\leq \ell_1<\ldots<\ell_r\leq N}
        \underset{a_p\neq a_q}{\sum_{a_1,\ldots,a_r=1}^r}
        \underset{b_p\neq b_q}{\sum_{b_1,\ldots,b_r=1}^r}
        \left[\underset{1\leq q<p\leq r}{\prod}\!\!\!\bar{s}_{m_{a_p} m_{a_q}} s_{n_{b_p}n_{b_q}}\right]
        \left[\overset{r}{\underset{p=1}{\prod}}\prod_{h=j_k}^{\ell_p-1}\bar{x}_{m_{a_p},h}x_{n_{b_p},h}\right]
        \nonumber
        \\
        &=
        \sum_{j_{k-1}\leq \ell_1<\ldots<\ell_r\leq N}
        \underset{a_p\neq a_q}{\sum_{a_1,\ldots,a_r=1}^r}\,
        \underset{b_p\neq b_q}{\sum_{b_1,\ldots,b_r=1}^r}
        \left[\underset{1\leq q<p\leq r}{\prod}\!\!\!\bar{s}_{m_{a_p} m_{a_q}} s_{n_{b_p}n_{b_q}}\right]
        \left[\overset{r}{\underset{p=1}{\prod}}\prod_{h=j_{k-1}}^{\ell_p-1}\bar{x}_{m_{a_p},h}x_{n_{b_p},h}\right]
        \nonumber
        \\
        &\times
        \Bigg[
            \overset{r}{\underset{q=1}{\prod}}
            \bar{x}_{m_p,j_k}x_{n_p,j_k} 
        \Bigg]\\
        &+\sum_{a_1,b_1=1}^r\sum_{j_{k-1}\leq \ell_2<\ldots<\ell_r\leq N}
        \underset{a_p\neq a_1}{\underset{a_p\neq a_q}{\sum_{a_2,\ldots,a_r=1}^r}}\,
        \underset{b_p\neq b_1}{\underset{b_p\neq b_q}{\sum_{b_2,\ldots,b_r=1}^r}}
        \left[\underset{2\leq q<p\leq r}{\prod}\!\!\!\bar{s}_{m_{a_p} m_{a_q}} s_{n_{b_p}n_{b_q}}\right]
        \left[\overset{r}{\underset{p=2}{\prod}}\prod_{h=j_{k-1}}^{\ell_p-1}\bar{x}_{m_{a_p},h}x_{n_{a_p},h}\right]
        \nonumber
        \\ 
        &\times
        \Bigg[
            \overset{r}{\underset{p=2}{\prod}}\, 
            \bar{s}_{m_{a_p} m_{a_1}}\bar{x}_{m_{a_p},j_k}
        \Bigg]
        \Bigg[
            \overset{r}{\underset{q=2}{\prod}}\, 
            s_{n_{a_q} n_{a_1}}x_{n_{a_q},j_k}
        \Bigg]
        \nonumber \ .
\end{align}

\subsection{Proof of Unitarity of Short Unitaries}

\label{appshortunitary}

To proof
(\ref{unitarityOmegajk}),
we write the matrix elements
\begin{equation}
    \begin{split}
    C_{k,\alpha\beta}^{[r]}
    =
    \sum_{i=0,1}
    \sum_{\lambda,\mu,\nu=1}^{D_i}
    \frac{\det B_{k-1,\alpha\rightarrow\lambda}^{[r-i]\dagger}}{\det B_{k-1}^{[r-i]\dagger}}
    \bar{B}_{k-1,\mu\lambda}^{[r-i]}
    B_{k-1,\mu\nu}^{[r-i]}
    \frac{\det B_{k-1,\nu\rightarrow\beta}^{[r-1]}}{\det B_{k-1}^{[r-i]}} 
    \end{split}
\end{equation}
where we introduced to alleviate notation 
\begin{equation}
    D_0=\binom{k-1}{r} \ , \quad D_1=\binom{k-1}{r-1} \ ,
\end{equation}
and we Cholesky-factorized the Gram matrix 
(\ref{Cab}) according to 
(\ref{CBB}).
Since
\begin{equation}
    \begin{split}
    &\frac{1}{\det B_{k-1}^{[r]}}
    \begin{vmatrix}
        \vphantom{x}_{k-1}\langle 1 | \Psi_{k-1,1}^{[r]} \rangle & \ldots & \vphantom{x}_{k-1}\langle 1 | \Psi_{k-1,D_0}^{[r]} \rangle & \vphantom{x}_{k}\langle 1 | \Psi_{k,\beta}^{[r]}\rangle  \\
        \ldots & \ldots & \ldots & \ldots \\
        \vphantom{x}_{k-1}\langle D_0 | \Psi_{k-1,1}^{[r]} \rangle & \ldots & \vphantom{x}_{k-1}\langle D_0 | \Psi_{k-1,D_0}^{[r]} \rangle & \vphantom{x}_{k}\langle D_0 | \Psi_{k,\beta}^{[r]}\rangle  \\
        \vphantom{x}_{k-1}\langle \mu | \Psi_{k-1,1}^{[r]} \rangle & \ldots & \vphantom{x}_{k-1}\langle \mu | \Psi_{k-1,D_0}^{[r]} \rangle & \vphantom{x}_{k}\langle \mu | \Psi_{k,\beta}^{[r]}\rangle  \\
    \end{vmatrix}\\
    =&~\vphantom{x}_{k}\langle \mu | \Psi_{k,\beta}^{[r]}\rangle - \sum_{\nu=1}^{D_0}
    B_{k-1,\mu\nu}^{[r]}
    \frac{\det B_{k-1,\nu\rightarrow\beta}^{[r]}}{\det B_{k-1}^{[r]}} = 0 \ ,
    \end{split}
\end{equation}
and
\begin{equation}
    \vphantom{x}_{k}\langle \mu+D_0 | \Psi_{k,\beta}^{[r]}\rangle 
    -
    \sum_{\nu=1}^{D_1}
    B_{k-1,\mu\nu}^{[r-1]}
    \frac{\det B_{k-1,\nu\rightarrow\beta}^{[r-1]}}{\det B_{k-1}^{[r-1]}} = 0 \ ,
\end{equation}
thanks to the fact the determinant of a matrix with repeated columns vanishes, 
we have
\begin{equation}
    \begin{split}
    C_{k,\alpha\beta}^{[r]}
    =
    \left[\sum_{\lambda=1}^{D_0}+\sum_{\lambda=D_0+1}^{D_0+D_1}\right]
    \langle\Psi_{k,\alpha}^{[r]}\ket{\lambda}\!\vphantom{x}_k\vphantom{x}_k\!\bra{\lambda}\Psi_{k,\beta}^{[r]}\rangle
    =
    \langle \Psi_{k,\alpha}^{[r]}|\Psi_{k,\beta}^{[r]}\rangle \ .
    \end{split}
\end{equation}
The proof of 
(\ref{unitarityOmegajk})  
is thus complete.

\setcounter{equation}{0}\renewcommand\theequation{A\arabic{equation}}

\bibliographystyle{bibliographicstyle}

\bibliography{ABCFbasis}

\providecommand{\href}[2]{#2}\begingroup\raggedright\begin{thebibliography}{10}

\bibitem{Caux10}
J.-S.~Caux, J.~Mossel, ``{Remarks on the notion of quantum integrability}'', \href{https://doi.org/10.1088/1742-5468/2011/02/P02023}{\emph{J. Stat. Mech.} {\bfseries 1102} (2011) P02023} [\href{https://arxiv.org/abs/1012.3587}{{\ttfamily 1012.3587}}].

\bibitem{Bethe31}
H.~Bethe, ``{Zur Theorie der Metalle. I. Eigenwerte und Eigenfunktionen der linearen Atomkette}'', \href{https://doi.org/10.1007/BF01341708}{\emph{Z. Phys.} {\bfseries 71} (1931) 205}.

\bibitem{Gaudin14}
M.~Gaudin, J.-S.~Caux~(Translator), \emph{The Bethe Wavefunction}, Cambridge University Press (2014).

\bibitem{Korepin93}
V.E.~Korepin, N.M.~Bogoliubov, A.G.~Izergin, \emph{Quantum Inverse Scattering Method and Correlation Functions}, Cambridge Monographs on Mathematical Physics, Cambridge University Press (1993), \href{https://doi.org/10.1017/CBO9780511628832}{10.1017/CBO9780511628832}.

\bibitem{Gomez96}
C.~G{\'o}mez, M.~Ruiz-Altaba, G.~Sierra, \emph{Quantum Groups in Two-Dimensional Physics}, Cambridge Monographs on Mathematical Physics, Cambridge University Press (1996), \href{https://doi.org/10.1017/CBO9780511628825}{10.1017/CBO9780511628825}.

\bibitem{Faddeev96}
L.D.~Faddeev, ``{How Algebraic Bethe Ansatz works for integrable model}'', \href{https://doi.org/10.48550/arXiv.hep-th/9605187}{\emph{{Les Houches School of Physics: Astrophysical Sources of Gravitational Radiation}} (1996) 149} [\href{https://arxiv.org/abs/hep-th/9605187}{{\ttfamily hep-th/9605187}}].

\bibitem{Wecker15}
D.~Wecker, M.B.~Hastings, N.~Wiebe, B.K.~Clark, C.~Nayak, M.~Troyer, ``Solving strongly correlated electron models on a quantum computer'', \href{https://doi.org/10.1103/physreva.92.062318}{\emph{Phys. Rev. A} {\bfseries 92} (2015) 1094} [\href{https://arxiv.org/abs/1506.05135}{{\ttfamily 1506.05135}}].

\bibitem{Kormos16}
M.~Kormos, M.~Collura, G.~Tak{\'a}cs, P.~Calabrese, ``{Real-time confinement following a quantum quench to a non-integrable model}'', \href{https://doi.org/10.1038/nphys3934}{\emph{Nat. Phys.} {\bfseries 13} (2016) 246} [\href{https://arxiv.org/abs/1604.03571}{{\ttfamily 1604.03571}}].

\bibitem{Robbiati24}
M.~Robbiati et~al., ``{Double-bracket quantum algorithms for high-fidelity ground state preparation}'',  \href{https://arxiv.org/abs/2408.03987}{{\ttfamily 2408.03987}}.

\bibitem{Nepomechie20}
R.I.~Nepomechie, ``{Bethe ansatz on a quantum computer?}'', \href{https://doi.org/10.26421/QIC21.3-4-4}{\emph{Quant. Inf. Comput.} {\bfseries 22} (2022) 0255} [\href{https://arxiv.org/abs/2010.01609}{{\ttfamily 2010.01609}}].

\bibitem{Raveh24ii}
D.~Raveh, R.I.~Nepomechie, ``{Estimating Bethe roots with VQE}'', \href{https://doi.org/10.1088/1751-8121/ad6db2}{\emph{J. Phys. A} {\bfseries 57} (2024) 355303} [\href{https://arxiv.org/abs/2404.18244}{{\ttfamily 2404.18244}}].

\bibitem{Crichigno24}
M.~Crichigno, A.~Prakash, ``{Quantum Spin Chains and Symmetric Functions}'',  \href{https://arxiv.org/abs/2404.04322}{{\ttfamily 2404.04322}}.

\bibitem{Faddeev81}
L.D.~Faddeev, L.A.~Takhtadzhyan, ``{Spectrum and Scattering of Excitations in the One-Dimensional Isotropic Heisenberg Model}'',  in \emph{Fifty Years of Mathematical Physics}, pp.~296--322 (1981), \href{https://doi.org/10.1142/9789814340960_0027}{DOI}.

\bibitem{VanDyke21}
J.S.~Van~Dyke, G.S.~Barron, N.J.~Mayhall, E.~Barnes, S.E.~Economou, ``{Preparing Bethe Ansatz Eigenstates on a Quantum Computer}'', \href{https://doi.org/10.1103/PRXQuantum.2.040329}{\emph{PRX Quantum} {\bfseries 2} (2021) 040329} [\href{https://arxiv.org/abs/2103.13388}{{\ttfamily 2103.13388}}].

\bibitem{VanDyke22}
J.S.~Van~Dyke, E.~Barnes, S.E.~Economou, R.I.~Nepomechie, ``{Preparing exact eigenstates of the open XXZ chain on a quantum computer}'', \href{https://doi.org/10.1088/1751-8121/ac4640}{\emph{J. Phys. A} {\bfseries 55} (2022) 055301} [\href{https://arxiv.org/abs/2109.05607}{{\ttfamily 2109.05607}}].

\bibitem{Raveh24}
D.~Raveh, R.I.~Nepomechie, ``{Deterministic Bethe state preparation}'', \href{https://doi.org/10.22331/q-2024-10-24-1510}{\emph{Quantum} {\bfseries 8} (2024) 1510} [\href{https://arxiv.org/abs/2403.03283}{{\ttfamily 2403.03283}}].

\bibitem{Li22}
W.~Li, M.~Okyay, R.I.~Nepomechie, ``{Bethe states on a quantum computer: success probability and correlation functions}'', \href{https://doi.org/10.1088/1751-8121/ac8255}{\emph{J. Phys. A} {\bfseries 55} (2022) 355305} [\href{https://arxiv.org/abs/2201.03021}{{\ttfamily 2201.03021}}].

\bibitem{Farias24}
R.M.S.~Farias, T.O.~Maciel, G.~Camilo, R.~Lin, S.~Ramos-Calderer, L.~Aolita, ``{Quantum encoder for fixed Hamming-weight subspaces}'',  \href{https://arxiv.org/abs/2405.20408}{{\ttfamily 2405.20408}}.

\bibitem{Sopena22}
A.~Sopena, M.H.~Gordon, D.~Garc{\'\i}a-Mart{\'\i}n, G.~Sierra, E.~L{\'o}pez, ``{Algebraic Bethe Circuits}'', \href{https://doi.org/10.22331/q-2022-09-08-796}{\emph{Quantum} {\bfseries 6} (2022) 796} [\href{https://arxiv.org/abs/2202.04673}{{\ttfamily 2202.04673}}].

\bibitem{Ruiz23}
R.~Ruiz, A.~Sopena, M.H.~Gordon, G.~Sierra, E.~L{\'o}pez, ``{The Bethe Ansatz as a Quantum Circuit}'', \href{https://doi.org/10.22331/q-2024-05-23-1356}{\emph{Quantum} {\bfseries 8} (2024) 1356} [\href{https://arxiv.org/abs/2309.14430}{{\ttfamily 2309.14430}}].

\bibitem{Murg12}
V.~Murg, V.E.~Korepin, F.~Verstraete, ``{Algebraic Bethe ansatz and tensor networks}'', \href{https://doi.org/10.1103/physrevb.86.045125}{\emph{Phys. Rev. B} {\bfseries 86} (2012) } [\href{https://arxiv.org/abs/1201.5627}{{\ttfamily 1201.5627}}].

\bibitem{Cirac20}
J.I.~Cirac, D.~P{\'e}rez-Garc{\'\i}a, N.~Schuch, F.~Verstraete, ``{Matrix product states and projected entangled pair states: Concepts, symmetries, theorems}'', \href{https://doi.org/10.1103/RevModPhys.93.045003}{\emph{Rev. Mod. Phys.} {\bfseries 93} (2021) 045003} [\href{https://arxiv.org/abs/2011.12127}{{\ttfamily 2011.12127}}].

\bibitem{Ruiz24}
R.~Ruiz, A.~Sopena, B.~Pozsgay, E.~L\'opez, ``{Efficient Eigenstate Preparation in an Integrable Model with Hilbert Space Fragmentation}'',  \href{https://arxiv.org/abs/2411.15132}{{\ttfamily 2411.15132}}.

\bibitem{Maillet96}
J.M.~Maillet, J.~Sanchez~de Santos, ``{Drinfel'd Twists and Algebraic Bethe Ansatz}'', {\emph{Am. Math. Soc. Trans. Ser. 2} {\bfseries 201} (2000) 137} [\href{https://arxiv.org/abs/q-alg/9612012}{{\ttfamily q-alg/9612012}}].

\bibitem{McAteer11}
S.G.~McAteer, M.~Wheeler, ``{Factorizing $F$-matrices and the XXZ spin-$\frac{1}{2}$ chain: A diagrammatic perspective}'', \href{https://doi.org/10.1016/j.nuclphysb.2011.05.019}{\emph{Nucl. Phys. B} {\bfseries 851} (2011) 346} [\href{https://arxiv.org/abs/1103.4488}{{\ttfamily 1103.4488}}].

\bibitem{Terras99}
V.~Terras, ``{Drinfel'd Twists and Functional Bethe Ansatz}'', \href{https://doi.org/10.1023/A:1007695001683}{\emph{{Lett. Math. Phys.}} {\bfseries 48} (1999) 263} [\href{https://arxiv.org/abs/math-ph/9902009}{{\ttfamily math-ph/9902009}}].

\bibitem{Pfeiffer00}
H.~Pfeiffer, ``{Factorizing twists and the universal $R$-matrix of the Yangian $Y(\mathfrak{sl}_{2})$}'', \href{https://doi.org/10.1088/0305-4470/33/48/323}{\emph{J. Phys. A} {\bfseries 33} (2000) 8929} [\href{https://arxiv.org/abs/math-ph/0006032}{{\ttfamily math-ph/0006032}}].

\bibitem{Pfeiffer00ii}
H.~Pfeiffer, ``{Factorizing twists and $R$-matrices for representations of the quantum affine algebra $U_q(\hat{\mathfrak{sl}}_2)$}'', \href{https://doi.org/10.1088/0305-4470/34/8/318}{\emph{J. Phys. A} {\bfseries 34} (2001) 1753} [\href{https://arxiv.org/abs/math-ph/0011029}{{\ttfamily math-ph/0011029}}].

\bibitem{Albert00}
T.D.~Albert, H.~Boos, R.~Flume, K.~Ruhlig, ``{Resolution of the nested hierarchy for rational $sl(n)$ models}'', \href{https://doi.org/10.1088/0305-4470/33/28/302}{\emph{J. Phys. A} {\bfseries 33} (2000) 4963} [\href{https://arxiv.org/abs/nlin/0002027}{{\ttfamily nlin/0002027}}].

\bibitem{Yang05}
W.-L.~Yang, Y.-Z.~Zhang, S.-Y.~Zhao, ``{Drinfeld twists and algebraic Bethe ansatz of the supersymmetric model associated with $U_q(gl(m|n))$}'', \href{https://doi.org/10.1007/s00220-005-1513-4}{\emph{Commun. Math. Phys.} {\bfseries 264} (2006) 87} [\href{https://arxiv.org/abs/hep-th/0503003}{{\ttfamily hep-th/0503003}}].

\bibitem{McAteer12}
S.G.~McAteer, M.~Wheeler, ``{On factorizing $F$-matrices in $\mathcal{Y}(sl_n)$ and $\mathcal{U}_q(\widehat{sl_n})$ spin chains}'', \href{https://doi.org/10.1088/1742-5468/2012/04/P04016}{\emph{J. Stat. Mech.} {\bfseries 1204} (2012) P04016} [\href{https://arxiv.org/abs/1112.0839}{{\ttfamily 1112.0839}}].

\bibitem{Albert00ii}
T.D.~Albert, H.~Boos, R.~Flume, R.H.~Poghossian, K.~Ruhlig, ``{The Drinfel'd twisted XYZ model}'', \href{https://doi.org/10.22323/1.006.0052}{\emph{PoS} {\bfseries tmr2000} (2000) 052} [\href{https://arxiv.org/abs/nlin/0005023}{{\ttfamily nlin/0005023}}].

\bibitem{Wang02}
Y.-S.~Wang, ``{The scalar products and the norm of Bethe eigenstates for the boundary XXX Heisenberg spin-1/2 finite chain}'', \href{https://doi.org/https://doi.org/10.1016/S0550-3213(01)00610-1}{\emph{Nucl. Phys. B} {\bfseries 622} (2002) 633}.

\bibitem{Kitanine07}
N.~Kitanine, K.K.~Kozlowski, J.M.~Maillet, G.~Niccoli, N.A.~Slavnov, V.~Terras, ``{Correlation functions of the open XXZ chain I}'', \href{https://doi.org/10.1088/1742-5468/2007/10/P10009}{\emph{J. Stat. Mech.} {\bfseries 0710} (2007) P10009} [\href{https://arxiv.org/abs/0707.1995}{{\ttfamily 0707.1995}}].

\bibitem{Yang04}
W.-L.~Yang, Y.-Z.~Zhang, S.-Y.~Zhao, ``{Drinfeld twists and algebraic Bethe ansatz of the supersymmetric $t$-$J$ model}'', \href{https://doi.org/10.1088/1126-6708/2004/12/038}{\emph{JHEP} {\bfseries 12} (2004) 038} [\href{https://arxiv.org/abs/cond-mat/0412182}{{\ttfamily cond-mat/0412182}}].

\bibitem{Zhao05}
S.-Y.~Zhao, W.-L.~Yang, Y.-Z.~Zhang, ``{Drinfeld twists and symmetric Bethe vectors of supersymmetric fermion models}'', \href{https://doi.org/10.1088/1742-5468/2005/04/P04005}{\emph{J. Stat. Mech.} {\bfseries 0504} (2005) P04005} [\href{https://arxiv.org/abs/nlin/0502050}{{\ttfamily nlin/0502050}}].

\bibitem{Maruyama10}
I.~Maruyama, H.~Katsura, ``{Continuous Matrix Product Ansatz for the One-Dimensional Bose Gas with Point Interaction}'', \href{https://doi.org/10.1143/JPSJ.79.073002}{\emph{J. Phys. Soc. Jap.} {\bfseries 79} (2010) 073002} [\href{https://arxiv.org/abs/1003.5463}{{\ttfamily 1003.5463}}].

\bibitem{Qiao18}
Y.~Qiao, X.~Zhang, K.~Hao, J.~Cao, G.-L.~Li, W.-L.~Yang et~al., ``{A convenient basis for the Izergin-Korepin model}'', \href{https://doi.org/https://doi.org/10.1016/j.nuclphysb.2018.03.010}{\emph{Nucl. Phys. B} {\bfseries 930} (2018) 399} [\href{https://arxiv.org/abs/1705.08114}{{\ttfamily 1705.08114}}].

\bibitem{Kitanine98}
N.~Kitanine, J.M.~Maillet, V.~Terras, ``{Form factors of the XXZ Heisenberg spin-$\frac 1 2$ finite chain}'', \href{https://doi.org/10.1016/S0550-3213(99)00295-3}{\emph{Nucl. Phys. B} {\bfseries 554} (1999) 647} [\href{https://arxiv.org/abs/math-ph/9807020}{{\ttfamily math-ph/9807020}}].

\bibitem{Martin11}
M.~Martins, M.~Zuparic, ``{The monodromy matrix in the F-basis for arbitrary six-vertex models}'', \href{https://doi.org/https://doi.org/10.1016/j.nuclphysb.2011.06.006}{\emph{Nucl. Phys. B} {\bfseries 851} (2011) 565} [\href{https://arxiv.org/abs/1106.0852}{{\ttfamily 1106.0852}}].

\bibitem{Zhong22}
C.~Zhong, ``{Computation of partition functions of free fermionic solvable lattice models via permutation graphs}'',  \href{https://arxiv.org/abs/2211.06805}{{\ttfamily 2211.06805}}.

\bibitem{Crampe15}
N.~Cramp{\'e}, ``{Algebraic Bethe ansatz for the totally asymmetric simple exclusion process with boundaries}'', \href{https://doi.org/10.1088/1751-8113/48/8/08FT01}{\emph{J. Phys. A} {\bfseries 48} (2015) 08FT01} [\href{https://arxiv.org/abs/1411.7954}{{\ttfamily 1411.7954}}].

\bibitem{Feher19}
G.~Feh{\'e}r, B.~Pozsgay, ``{The propagator of the finite XXZ spin-$\frac{1}{2}$ chain}'', \href{https://doi.org/10.21468/scipostphys.6.5.063}{\emph{SciPost Phys.} {\bfseries 6} (2019) 099} [\href{https://arxiv.org/abs/1808.06279}{{\ttfamily 1808.06279}}].

\bibitem{Katsura12}
H.~Katsura, I.~Maruyama, ``{Matrix Product States in Quantum Integrable Models}'',  in \emph{Frontiers in Quantum Information Research}, pp.~302--321 (2012), \href{https://doi.org/10.1142/9789814407199_0012}{DOI}.

\bibitem{Alcaraz03}
F.C.~Alcaraz, M.J.~Lazo, ``{The Bethe ansatz as a matrix product ansatz}'', \href{https://doi.org/10.1088/0305-4470/37/1/L01}{\emph{J. Phys. A} {\bfseries 37} (2004) L1} [\href{https://arxiv.org/abs/cond-mat/0304170}{{\ttfamily cond-mat/0304170}}].

\bibitem{Alcaraz03ii}
F.C.~Alcaraz, M.J.~Lazo, ``{Exact solutions of exactly integrable quantum chains by a matrix product ansatz}'', \href{https://doi.org/10.1088/0305-4470/37/14/001}{\emph{J. Phys. A} {\bfseries 37} (2004) 4149} [\href{https://arxiv.org/abs/cond-mat/0312373}{{\ttfamily cond-mat/0312373}}].

\bibitem{Alcaraz06}
F.C.~Alcaraz, M.J.~Lazo, ``{Generalization of the matrix product ansatz for integrable chains}'', \href{https://doi.org/10.1088/0305-4470/39/36/014}{\emph{J. Phys. A} {\bfseries 39} (2006) 11335} [\href{https://arxiv.org/abs/cond-mat/0608177}{{\ttfamily cond-mat/0608177}}].

\bibitem{Katsura10}
H.~Katsura, I.~Maruyama, ``{Derivation of the matrix product ansatz for the Heisenberg chain from the algebraic Bethe ansatz}'', \href{https://doi.org/10.1088/1751-8113/43/17/175003}{\emph{J. Phys. A} {\bfseries 43} (2010) 175003} [\href{https://arxiv.org/abs/0911.4215}{{\ttfamily 0911.4215}}].

\bibitem{Ovchinnikov10}
A.A.~Ovchinnikov, ``{Coordinate space wave function from the Algebraic Bethe Ansatz for the inhomogeneous six-vertex model}'', \href{https://doi.org/10.1016/j.physleta.2010.01.022}{\emph{Phys. Lett. A} {\bfseries 374} (2010) 1311} [\href{https://arxiv.org/abs/1001.2672}{{\ttfamily 1001.2672}}].

\bibitem{Zamolodchikov79}
A.B.~Zamolodchikov, A.B.~Zamolodchikov, ``{Factorized S-matrices in two dimensions as the exact solutions of certain relativistic quantum field theory models}'', \href{https://doi.org/https://doi.org/10.1016/0003-4916(79)90391-9}{\emph{Annals Phys.} {\bfseries 120} (1979) 253}.

\bibitem{Ovchinnikov00}
A.A.~Ovchinnikov, ``{Construction of monodromy matrix in the F-basis and scalar products in spin chains}'', \href{https://doi.org/10.1142/S0217751X01003743}{\emph{Int. J. Mod. Phys. A} {\bfseries 16} (2001) 2175} [\href{https://arxiv.org/abs/math-ph/0012007}{{\ttfamily math-ph/0012007}}].

\bibitem{Takahashi99}
M.~Takahashi, \emph{{Thermodynamics of One-Dimensional Solvable Models}}, Cambridge University Press (1999), \href{https://doi.org/10.1017/cbo9780511524332}{10.1017/cbo9780511524332}.

\bibitem{vanTogeren16}
S.J.~van Tongeren, ``{Introduction to the thermodynamic Bethe ansatz}'', \href{https://doi.org/10.1088/1751-8113/49/32/323005}{\emph{J. Phys. A} {\bfseries 49} (2016) 323005} [\href{https://arxiv.org/abs/1606.02951}{{\ttfamily 1606.02951}}].

\bibitem{Frahm13}
H.~{Frahm}, A.~{Seel}, ``{The staggered six-vertex model: Conformal invariance and corrections to scaling}'', \href{https://doi.org/10.1016/j.nuclphysb.2013.12.015}{\emph{Nucl. Phys. B} {\bfseries 879} (2014) 382} [\href{https://arxiv.org/abs/1311.6911}{{\ttfamily 1311.6911}}].

\bibitem{Vanicat17}
M.~{Vanicat}, L.~{Zadnik}, T.~{Prosen}, ``{Integrable Trotterization: Local Conservation Laws and Boundary Driving}'', \href{https://doi.org/10.1103/PhysRevLett.121.030606}{\emph{Phys. Rev. Lett.} {\bfseries 121} (2018) 030606} [\href{https://arxiv.org/abs/1712.00431}{{\ttfamily 1712.00431}}].

\bibitem{Richardson77}
R.W.~Richardson, ``Pairing in the limit of a large number of particles'', \href{https://doi.org/10.1063/1.523493}{\emph{J. Math. Phys.} {\bfseries 18} (1977) 1802}.

\bibitem{Cambiaggio97}
M.C.~Cambiaggio, A.M.F.~Rivas, M.~Saraceno, ``{Integrability of the pairing hamiltonian}'', \href{https://doi.org/10.1016/S0375-9474(97)00418-1}{\emph{Nucl. Phys. A} {\bfseries 624} (1997) 157} [\href{https://arxiv.org/abs/nucl-th/9708031}{{\ttfamily nucl-th/9708031}}].

\bibitem{Amico00}
L.~Amico, G.~Falci, R.~Fazio, ``{The BCS model and the off-shell Bethe ansatz for vertex models}'', \href{https://doi.org/10.1088/0305-4470/34/33/307}{\emph{J. Phys. A} {\bfseries 34} (2001) 6425} [\href{https://arxiv.org/abs/cond-mat/0010349}{{\ttfamily cond-mat/0010349}}].

\bibitem{Amico01}
L.~Amico, A.~Di~Lorenzo, A.~Osterloh, ``{Integrable models for confined fermions: applications to metallic grains}'', \href{https://doi.org/10.1016/S0550-3213(01)00385-6}{\emph{Nucl. Phys. B} {\bfseries 614} (2001) 449} [\href{https://arxiv.org/abs/cond-mat/0105537}{{\ttfamily cond-mat/0105537}}].

\bibitem{vonDelft02}
J.~von Delft, R.~Poghossian, ``{Algebraic Bethe ansatz for a discrete-state BCS pairing model}'', \href{https://doi.org/10.1103/physrevb.66.134502}{\emph{Phys. Rev. B} {\bfseries 66} (2002) } [\href{https://arxiv.org/abs/cond-mat/0106405}{{\ttfamily cond-mat/0106405}}].

\bibitem{workinprogress}
E.~L{\'o}pez, B.~Pozsgay, R.~Ruiz, G.~Sierra, A.~Sopena, ``{work in progress}.''

\bibitem{Baxter72}
R.J.~Baxter, ``{Eight vertex model in lattice statistics and one-dimensional anisotropic Heisenberg chain. I. Some fundamental eigenvectors}'', \href{https://doi.org/10.1016/0003-4916(73)90439-9}{\emph{Annals Phys.} {\bfseries 76} (1973) 1}.

\bibitem{Takhtadzhan79}
L.A.~Takhtadzhan, L.D.~Faddeev, ``{The Quantum Method of the Inverse Problem and the Heisenberg XYZ Model}'', \href{https://doi.org/10.1070/RM1979v034n05ABEH003909}{\emph{Russ. Math. Surveys} {\bfseries 34} (1979) 11}.

\bibitem{Felder94}
G.~Felder, ``{Elliptic quantum groups}'',  in \emph{{11$\vphantom{x}^{th}$ International Conference on Mathematical Physics (ICMP-11) \textnormal{(Satellite colloquia: New Problems in the General Theory of Fields and Particles, Paris, France, July 25–28, 1994)}}}, pp.~211--218, 7, 1994 [\href{https://arxiv.org/abs/hep-th/9412207}{{\ttfamily hep-th/9412207}}].

\bibitem{Gromov16}
N.~Gromov, F.~Levkovich-Maslyuk, G.~Sizov, ``{New construction of eigenstates and separation of variables for $\mathrm{SU}(N)$ quantum spin chains}'', \href{https://doi.org/10.1007/JHEP09(2017)111}{\emph{JHEP} {\bfseries 2017} (2017) 111} [\href{https://arxiv.org/abs/1610.08032}{{\ttfamily 1610.08032}}].

\bibitem{Liashyk18}
A.~{Liashyk}, N.A.~{Slavnov}, ``{On Bethe vectors in $\mathfrak{gl}_3$-invariant integrable models}'', \href{https://doi.org/10.1007/JHEP06(2018)018}{\emph{JHEP} {\bfseries 2018} (2018) 18} [\href{https://arxiv.org/abs/1803.07628}{{\ttfamily 1803.07628}}].

\end{thebibliography}\endgroup

\end{document}